\documentclass[12pt]{spieman} 
\usepackage{amsmath,amsfonts,amssymb}
\usepackage{graphicx}
\usepackage{setspace}
\usepackage{tocloft}
\usepackage{lineno}
\usepackage{pifont}
\usepackage{booktabs} 
\usepackage{epsfig}
\usepackage{amssymb}
\usepackage{amsmath}
\usepackage{wrapfig}
\usepackage{graphicx}
\usepackage{array}
\usepackage{booktabs}
\usepackage{xcolor}
\usepackage{multirow}
\usepackage{pifont}
\usepackage{textgreek}
\usepackage{tabularx}
\usepackage{array}
\usepackage{caption}
\usepackage{booktabs}
\usepackage{multirow}
\usepackage{bm}
\usepackage{graphicx}
\usepackage{threeparttable}
\usepackage[margin=1in]{geometry}
\newcommand{\std}[1]{\scriptsize$\pm$#1}

\title{Benchmarking Deep Learning-Based Reconstruction Methods for Photoacoustic Computed Tomography with Clinically Relevant Synthetic Datasets}

\author[a]{Panpan Chen}
\author[a]{Seonyeong Park}
\author[a]{Gangwon Jeong}
\author[b]{Refik Mert Cam}
\author[c,d]{Umberto Villa}
\author[a,b*]{Mark A. Anastasio}

\affil[a]{Department of Bioengineering, University of Illinois Urbana-Champaign, Urbana, IL 61801, USA}
\affil[b]{Department of Electrical and Computer Engineering, University of Illinois Urbana-Champaign, Urbana, IL 61801, USA}
\affil[c]{Department of Biomedical Engineering, The University of Texas at Austin, Austin, TX 78712, USA}
\affil[d]{Oden Institute for Computational Engineering and Sciences, The University of Texas at Austin, Austin, TX 78712, USA}

\cftpagenumbersoff{figure}
\cftpagenumbersoff{table} 
\begin{document} 
\maketitle

\begin{abstract} \\
\textbf{Purpose:} Deep learning (DL)–based image reconstruction methods for photoacoustic computed tomography (PACT) have developed rapidly in recent years. However, most existing methods have not employed standardized datasets, and their evaluations rely on traditional image quality (IQ) metrics that may lack clinical relevance. The absence of a standardized framework for clinically meaningful IQ assessment hinders fair comparison and raises concerns about the reproducibility and reliability of reported advancements in PACT. \\
\textbf{Approach:} A benchmarking framework is proposed that provides open-source, anatomically plausible synthetic datasets and evaluation strategies for DL-based acoustic inversion methods in PACT. The datasets each include over 11,000 two-dimensional (2D) stochastic breast objects with clinically relevant lesions and paired measurements at varying modeling complexity. The evaluation strategies incorporate both traditional and task-based IQ measures to assess fidelity and clinical utility. A preliminary benchmarking study is conducted to demonstrate the framework's utility by comparing DL-based and physics-based reconstruction methods. \\
\textbf{Results:} The benchmarking study demonstrated that the proposed framework enabled comprehensive, quantitative comparisons of reconstruction performance and revealed important limitations in certain DL-based methods. Although they performed well according to traditional IQ measures, they often failed to accurately recover lesions. This highlights the inadequacy of traditional metrics and motivates the need for task-based assessments.\\
\textbf{Conclusion:} The proposed benchmarking framework enables systematic comparisons of DL–based acoustic inversion methods for 2D PACT. By integrating clinically relevant synthetic datasets with rigorous evaluation protocols, it enables reproducible, objective assessments and facilitates method development and system optimization in PACT.
\end{abstract}


% Include email contact information for corresponding author
{\noindent \footnotesize\textbf{*}Mark A. Anastasio,  \linkable{maa@illinois.edu} }

\begin{spacing}{2} 

\section{Introduction}
\label{introduction}
Photoacoustic computed tomography (PACT), also known as optoacoustic tomography, is an emerging imaging modality that exploits optical contrast mechanisms and ultrasound detection principles \cite{wang2012photoacoustic}. In PACT, biological tissues are irradiated with a short laser %-
pulse, %d lasers, 
resulting in the absorption of optical energy within the tissues. Under thermal confinement conditions, the absorbed optical energy induces the generation of acoustic pressure wavefields via the photoacoustic effect \cite{xia2014photoacoustic}. These pressure waves propagate outward and are subsequently detected by broadband ultrasonic transducers positioned along the measurement surface. From these measurements, tomographic reconstruction methods can be employed to estimate the spatially varying initial pressure distribution \cite{poudel2019survey}, which is proportional to the deposited optical energy density. The hybrid nature of PACT, which combines optical contrast with ultrasonic detection, provides excellent endogenous optical absorption contrast at high resolution, establishing it as a promising technology for diverse anatomical, functional, and molecular imaging applications in both preclinical research and clinical medicine \cite{yao2016multiscale}. 

Various physics-based reconstruction methods have been proposed for PACT \cite{xu2005universal, kunyansky2007explicit, wang2012simple, buehler2011model, jeong2025revisiting}. However, many of these %PACT reconstruction 
approaches face significant challenges in practical imaging scenarios. One challenge involves reconstructing images from incomplete measurements, typically referred to as limited-view and sparse-view problems \cite{xu2004reconstructions, frikel2015artifacts}. Such conditions lead to inaccurate reconstructions with artifacts. Limited-view problems arise when the angular coverage of the detection aperture is restricted, whereas sparse-view problems occur when spatial sampling is insufficient.  
Furthermore, conventional reconstruction algorithms often assume %ideal 
point-like transducers, neglecting %; however, 
finite transducer aperture sizes and constrained scanning geometries. This introduces spatially variant system responses that deviate from the assumed linear shift-invariant model, causing resolution degradation and geometric distortions \cite{mitsuhashi2014investigation}. Moreover, acoustic heterogeneity presents another challenge. Because biological tissues exhibit spatially varying %heterogeneous 
acoustic properties with frequency-dependent attenuation, methods based on the assumption of %that assume 
an acoustically homogeneous, lossless medium fail to account for these variations and consequently can produce %severe 
distortions that become severe when acoustic heterogeneity is pronounced \cite{poudel2019survey}. Although %While 
recent model-based methods incorporating more accurate physical modeling have been proposed to compensate for these limitations, such approaches are generally computationally intensive \cite{pattyn2021model, matthews2018parameterized, jeong2025revisiting}.

Recently, in addition to the well-established physics-based approaches, deep learning (DL)-based image reconstruction methods have been actively developed for PACT \cite{grohl2021deep, hauptmann2020deep}. As data-driven approaches, DL methods can leverage learned data-driven priors to model complex mappings to enhance image reconstruction quality under non-ideal conditions and enable near real-time imaging through rapid inference \cite{hauptmann2018model, schwab2019learned, jeon2021deep, chen2025learning}. These methods can be broadly categorized into supervised, semi-supervised, and unsupervised approaches, which differ primarily in their dependence on labeled data \cite{hauptmann2020deep}. The performance of these paradigms depends critically on large, diverse, and representative training datasets that capture application-specific variability to characterize the statistical distribution of the target domain \cite{grohl2021deep}. For instance, in computer vision, benchmark datasets such as ImageNet \cite{deng2009imagenet} have driven rapid progress by providing large-scale, publicly available data for model training and performance comparison. However, only a few open-source datasets applicable to PACT are currently available \textcolor{blue}{\cite{davoudi2019deep, bench2020toward, ozdemir2022oadat}}.

To effectively support the development of artificial intelligence %development 
in PACT, a dataset comprised of well-characterized (to-be-imaged) objects and their corresponding tomographic data should meet the following requirements: (1) objects should exhibit realistic anatomical variability; (2) objects should represent plausible physiological, acoustic, and optical property variations; (3) objects should incorporate clinically relevant pathological features, such as lesions, to enable task-based image quality (IQ) assessments; and (4) the measurement data shall be acquired using a specified imaging system configuration and must embody the underlying imaging physics, including inherent uncertainties and errors.

Experimental datasets of raw PACT measurements have been considered for benchmarking purposes \cite{davoudi2019deep, huang2021functional}. For example, Ozdemir et al. \cite{ozdemir2022oadat} recently introduced a large-scale forearm dataset comprised of more than 70,000 \textit{in vivo} photoacoustic measurements acquired under various imaging configurations. However, it is often infeasible %difficult 
to obtain ground truth from \textit{in vivo} experimental datasets, which is essential for supervised learning \cite{dispirito2021sounding}. Moreover, the collection of large-scale clinical measurement data is difficult, particularly for emerging modalities such as PACT. While well-characterized physical phantoms can provide ground truth data, the construction and imaging of a large and diverse ensemble of experimental phantoms is time-consuming and costly \cite{christie2023review, hacker2024tutorial}.

In contrast, synthetic PACT datasets comprised of paired objects and corresponding simulated measurements offer a flexible and cost-effective alternative \cite{lou2017generation, li20213}. Through stochastic generation of numerical phantoms that are virtually imaged using high-fidelity multiphysics modeling of the PACT forward process, large-scale ensembles of objects and corresponding measurement data can be generated readily with %while providing 
known ground truth by design. Such approaches afford complete control over anatomical, physiological, and pathological variability while maintaining realistic tissue property distributions and incorporating disease-specific features \cite{park2023stochastic}. The resulting well-characterized datasets serve as a solid basis for comprehensive benchmarking.

This study proposes a benchmarking framework for PACT image reconstruction that integrates open-source synthetic data and objective IQ assessment methodologies based on lesion detection tasks. Motivated by a 2D ring-array imaging system \cite{lin2018single}, two-dimensional (2D) PACT is considered in this work. However, the framework may also find application for early-stage prototyping of methods that can eventually be extended to use with 3D PACT systems \cite{hauptmann2020deep, wang2023adaptive, wang2023photoacoustic}. The synthetic datasets comprise realistic 2D breast object representations derived from previously developed three-dimensional (3D) numerical breast phantoms (NBPs) \cite{park2023stochastic}, which reflect clinically relevant anatomical variations and photoacoustic tissue properties. A high-fidelity virtual imaging framework is implemented to simulate photoacoustic measurements. The evaluation strategies integrate both traditional and task-based measures of IQ, facilitating the assessment of physical and statistical properties of reconstructed images along with their utility for diagnostic tasks. The proposed datasets and code packages for the evaluation strategies are publicly available \cite{data, code}.

The remainder of the paper is organized as follows. In Section \ref{background}, %2, 
the PACT imaging model, the motivation for virtual imaging studies, %trails, 
and the salient aspects of task-based IQ assessment are reviewed. The proposed benchmarking framework is detailed %presented 
in Section \ref{method}. %3. 
%The e
Example benchmarking studies that demonstrate the utility of the proposed benchmarking framework are described in Section \ref{benchmarking_study}, %4, 
with the corresponding results provided in Section \ref{benchmarking_study_result}. %5. 
Finally, the paper concludes with a discussion in Section \ref{conclusion}. %6.

\section{Background}
\label{background}
This section reviews the PACT imaging model, virtual imaging studies for PACT, and task-based IQ assessment methods.

\subsection{Photoacoustic Computed Tomography }
In PACT, the sought-after object is irradiated by a short laser pulse, resulting in optical energy absorption denoted by $A(\bm{r})$, where $\bm{r} \in \mathbb{R}^3$ represents the spatial position. Through the photoacoustic effect, this absorbed energy induces a localized acoustic pressure rise, generating an initial pressure distribution $p_0(\bm{r})$ = $\Gamma(\bm{r}) A(\bm{r})$, where $\Gamma(\bm{r})$ is the dimensionless Gr\"uneisen parameter. The resulting initial pressure distribution subsequently propagates as acoustic pressure waves $p(\bm{r}, t)$ through the object and surrounding medium. These waves are recorded by ultrasonic transducers positioned on a measurement aperture, %$\Omega_0 \subset \mathbb{R}^3$, 
yielding the pressure measured at the transducer locations. The detected pressure is recorded over the time interval $t \in [0, T]$, where $t=0$ denotes the laser excitation time and $T$ is the end of the acquisition period. %signals.
%$p(\bm{r}_0,t)$. Here, $\bm{r}_0 \in \mathbb{R}^3$ % \in \Omega_0$ denotes the transducer location, and $t \in [0, T]$ represents the acquisition time, %frame, 
%with $t=0$ denoting the time when the tissue was excited by the laser pulse and $T$ being the end of the acquisition period.

For lossy, acoustically heterogeneous media, the propagation of $p(\bm{r},t)$ can be described %is governed
by the following coupled system of first-order partial differential equations \cite{huang2013full}:
\begin{equation}
\label{eq:wave-equation}
\begin{aligned}
\frac{\partial}{\partial t} \bm{u}(\bm{r}, t) &= -\frac{1}{\rho_0(\bm{r})} \nabla p(\bm{r}, t), \\
\frac{\partial}{\partial t} \rho(\bm{r}, t) &= -\rho_0(\bm{r}) \nabla \cdot \bm{u}(\bm{r}, t), \\
p(\bm{r}, t) &= c_0(\bm{r})^2 \left\{ 1 - \mu(\bm{r}) \frac{\partial}{\partial t}(-\nabla^2)^{y/2 - 1} - \eta(\bm{r})(-\nabla^2)^{(y - 1)/2} \right\} \rho(\bm{r}, t), %.
\end{aligned}
\end{equation}
subject to the initial conditions:
\begin{equation}
\label{eq:initial-condition}
\left.p_0(\bm{r}) \equiv p(\bm{r}, t)\right|_{t=0}=\Gamma(\bm{r}) A(\bm{r}),\left.\quad \bm{u}(\bm{r}, t)\right|_{t=0}=\bm{0}.
\end{equation}
Here, $\bm{u}(\bm{r}, t)$ represents the acoustic particle velocity, $c_0(\bm{r})$ denotes the speed-of-sound (SOS) distribution, and $\rho(\bm{r})$ and $\rho_0(\bm{r})$ represent the medium's acoustic density and ambient density distributions, respectively. 
The quantities $\mu(\bm{r})$ and $\eta(\bm{r})$ describe the acoustic absorption and dispersion proportionality coefficients:
\begin{equation}
    \label{eq:coeff-final}
    \mu(\bm{r}) = -2 \alpha_0(\bm{r}) c_0(\bm{r})^{y - 1}, \quad \eta(\bm{r}) = 2 \alpha_0(\bm{r}) c_0(\bm{r})^{y} \tan\left( \frac{\pi y}{2} \right), 
\end{equation} 
where $\alpha_0(\bm{r})$ is the frequency-independent attenuation coefficient and $y$ is the power law exponent.

From the measured acoustic data, %$p(\bm{r}_0,t)$, 
the initial pressure distribution $p_0(\bm{r})$ can be reconstructed by solving an acoustic inverse problem~\cite{poudel2019survey}. Numerous reconstruction methods have been developed for PACT, which can be broadly categorized into physics-based and data-driven approaches \cite{poudel2019survey, hauptmann2020deep}. Physics-based methods include analytical methods, non-iterative numerical methods (such as time-reversal algorithms), and iterative optimization-based methods \cite{xu2005universal, xu2004time, buehler2011model}, all of which leverage %ing 
established imaging physics to achieve robust reconstruction. However, analytical methods often exhibit %suffer from 
degraded performance in non-ideal scenarios, whereas iterative methods require high computational cost. Alternatively, DL-based reconstruction methods adopt a data-driven paradigm to learn the inverse mappings without explicit physical modeling \cite{jeon2020deep,chen2020improved,wang2023adaptive, chen2025learning}. While such DL-based methods demonstrate potential for achieving both computational efficiency and reconstruction accuracy in complex scenarios, their stability and generalizability remain open challenges.

\subsection{Virtual Imaging Studies for PACT}
Systematic and rigorous evaluation of PACT reconstruction methods is essential for their development and optimization toward clinical translation. However, directly assessing algorithm performance in experimental studies is challenging, as the experimental data generally do not provide ideal ground truth for quantitative evaluation of reconstruction performance. %Specifically
Particularly, \textit{in vivo} studies pose challenges in determining ground truth and are constrained by ethical considerations. Although physical phantoms can offer known ground truth, their construction is often costly, and factors such as system calibration errors and measurement noise can introduce uncertainties that limit their reliability. In contrast,
% \textcolor{cyan}{However, directly assessing algorithm performance in experimental PACT studies is challenging, as IQ is affected not only by the reconstruction algorithm itself but also by inter-subject variability and hardware-specific factors such as transducer characteristics, data acquisition settings, and system noise.}
% \textcolor{cyan}{Additionally, experimental studies} \textcolor{red}{involve high costs and ethical limitations, which hinder iterative optimization of reconstruction algorithms during early-stage development.} 
virtual imaging (VI) studies provide a cost-effective and controlled alternative for systematic evaluation of PACT reconstruction methods. By employing synthetic objects and simulated measurement data, VI studies allow precise control of parameters to assess and refine methods. In addition, because the ground truth is explicitly defined, VI studies support quantitative and %evaluation of image quality and facilitate 
objective assessments of IQ %image quality 
that align algorithm development with clinical requirements.

%In PACT, VI studies emulate the imaging pipeline by integrating 
A VI study pipeline for PACT integrates numerical phantoms, PACT forward process modeling for the target imaging system, and quantitative IQ evaluation. % metrics. 
Numerical phantoms, constructed through mathematical or computational modeling of anatomical structures with tissue-specific physiological and photoacoustic properties, provide the ground truth. To ensure clinical relevance, numerical phantoms must realistically capture stochastic anatomical and physiological variations, including both intrinsic tissue characteristics and pathology-associated alterations \cite{li20213, park2023stochastic}. In PACT, several %such numerical phantoms such as 
NBPs have been developed % for specific applications 
\cite{lou2017generation, park2023stochastic, cam2025numerical, choe2009differentiation}, which correspond to representative examples of such numerical phantoms .
% stochastic, anatomically realistic breast structures and stochastically represent anatomical and physiological variations,

Virtual PACT imaging involves a two-stage physics-based modeling framework that incorporates both optical and acoustic processes \cite{xia2014photoacoustic}. In the optical stage, photon transport through biological tissues is simulated to compute the spatial distribution of absorbed optical energy $A(\bm{r})$. Methods such as Monte Carlo techniques (e.g., MCX \cite{fang2009monte}) or diffusion approximations \cite{tarvainen2012reconstructing} account for wavelength-dependent optical properties and realistic illumination conditions to generate the initial pressure distribution $p_0(\bm{r})$. %for photoacoustic signal generation. 
In the acoustic stage, acoustic wave propagation is simulated to obtain the %received 
pressure measurements acquired by transducers $p(\bm{r}, t)$ according to Equations. \eqref{eq:wave-equation} and \eqref{eq:initial-condition}, while considering acoustic medium properties, detection geometry, and transducer response. Established tools such as k-Wave \cite{treeby2010k} can be used for %to generate 
customizable acoustic simulations that closely replicate experimental conditions.

\subsection{Task-based Image Quality Assessment}
\label{sec:2-3}
Quantitative evaluation plays a central role in evaluating and optimizing reconstruction algorithms for PACT. Physical measures of IQ have been widely employed in PACT imaging, including spatial resolution, contrast-to-noise ratio (CNR) \cite{mitcham2015photoacoustic, najafzadeh2020photoacoustic}, structural similarity index (SSIM) \cite{yang2023recent, awasthi2020deep}, and mean squared error (MSE) \cite{yang2023recent}. These traditional metrics provide standardized benchmarks for assessing image fidelity, noise characteristics, and structural preservation in reconstructed images. However, while these physical IQ measures effectively quantify certain image properties, they do not always correlate with the utility of images with consideration of a specified diagnostic task such as lesion detection. This limitation motivates the adoption of task-based measures of IQ, which are introduced in the following section.

Task-based IQ assessment provides a rigorous framework for developing and optimizing PACT reconstruction methods by directly quantifying image utility for specific clinical objectives \cite{lou2016application, petschke2013comparison}. In contrast to traditional IQ metrics, task-based measures of IQ quantify the utility of an image for a specified clinical purpose based on observer performance. \cite{eckstein2001model, jha2021objective}. Here, the observer refers to the entity that performs the clinically relevant task. While the observer can be a human reader, early stage assessments via VI studies typically employ numerical observer models. These are computer algorithms to enable objective, reproducible, and computationally efficient evaluations \cite{li2023estimating, li2024application}. Clinical tasks can be broadly categorized into three types: estimation tasks, where the objective is to measure a parameter of interest, classification tasks, where the goal is to discriminate among multiple possible states (e.g., presence or absence of a lesion), and hybrid tasks that combine multiple objectives (e.g., detection-estimation). This work focuses exclusively on classification tasks.

In classification tasks, the observer must make a decision among multiple hypotheses $\{H_0, H_1,$ $\ldots, H_{n_C - 1}\}$, where each hypothesis $H_i$ represents a distinct clinical condition and $n_C$ denotes the total number of classes. For example, in a binary lesion detection task, the observer chooses between signal-present ($H_1$) and signal-absent ($H_0$) hypotheses. The observer makes decisions by first computing a test statistic $t_i(\mathbf{g})$ that maps the observed data vector $\mathbf{g}$ (i.e., reconstructed images or measurement data) to a real-valued scalar variable. The observer then applies a %the 
decision rule to select the hypothesis with the maximum test statistic \cite{yendiki2005analysis}:
\begin{equation}
\text{Decide } H_i \text{ if } t_i(\mathbf{g}) > t_j(\mathbf{g}), 
\quad \forall j \neq i.
\end{equation}
In the special case of binary classification (i.e., $n_C = 2$), the decision rule simplifies to a comparison between %comparing 
a single test statistic $t=t(\mathbf{g})$ and %to 
a decision threshold $\tau$:
\begin{equation}
\text{Decide } H_1 \text{ if } t(\mathbf{g}) > \tau; 
\quad \text{otherwise decide } H_0.
\end{equation}

Various numerical observers have been developed for objective IQ assessment \cite{eckstein2001model, vaishnav2014objective}. The ideal observer\cite{li2023estimating} is a specific type of numerical observer that implements an optimal decision strategy. However, this approach requires complete knowledge of image statistics and is generally difficult to compute.
% While the ideal observer achieves optimal performance by utilizing complete statistical information, it is often computationally intractable for complex tasks. 
Linear observers offer practical alternatives, with the Hotelling observer (HO) serving as the optimal linear discriminant based on first- and second-order statistical moments \cite{gallas2003validating}. As an extension to HO, the channelized Hotelling observer (CHO) incorporates channels to reduce data dimensionality while preserving task-relevant information. Depending on the task, different %Various 
figures of merit (FOMs) can be employed %depending on the task 
to evaluate the observer performance \cite{plativsa2011channelized}. For binary detection tasks, commonly used measures of signal detection or discrimination performance include the signal-to-noise ratio of the test statistic (SNR$_t$) and the area under the receiver operating characteristic curve (AUC).

\section{Proposed Benchmarking Framework}
\label{method}
This section introduces the proposed benchmarking pipeline for assessing PACT image reconstruction methods, comprising clinically relevant synthetic breast datasets \cite{chen2025benchmarking} and evaluation strategies that integrate both physical and task-based measures of IQ.

\subsection{Proposed Synthetic Datasets for Breast PACT} 
As shown in Table \ref{tab:datasets_overview}, three synthetic datasets for 2D breast PACT were generated with progressively increasing acoustic modeling complexity. Each dataset consists of 2D stochastic breast objects and corresponding measurements. Dataset 1 assumes a homogeneous acoustic medium with point-like transducers; Dataset 2 incorporates acoustic heterogeneity while retaining point-like transducers; and Dataset 3 employs both acoustic heterogeneity and finite-sized transducers. For data generation, previously developed stochastic 3D NBP data \cite{park2023stochastic} were utilized, which included corresponding initial pressure distributions obtained from 3D photon transport simulations \cite{park2025virtual}. In this study, 2D objects were derived from the 3D NBP data, and subsequently, the acoustic measurement acquisition process was modeled in 2D. 
%The data generation employed previously developed stochastic 3D NBPs and a virtual imaging framework \cite{park2023stochastic}, in which photon transport was simulated in 3D and acoustic measurement acquisition was modeled in 2D.
\begin{table}[h]
    \centering
    \caption{Summary of the Generated Breast Datasets for PACT}
    \label{tab:datasets_overview} 
    \begin{threeparttable}
    \resizebox{0.99\linewidth}{!}{%
    \begin{tabular}{
        p{2cm} p{1.5cm} p{2.5cm} 
        p{1cm} p{3.5cm} p{3.5 cm} p{4.0 cm} 
        p{3.5 cm}
    }
        \toprule
        \multirow{3}{*}{\textbf{Dataset}} 
        & \multirow{3}{*}{\textbf{Size}} 
        & \multirow{3}{*}{\begin{tabular}[c]{@{}l@{}}\textbf{Acoustic}\\ \textbf{Heterogeneity} \end{tabular}} 
        & \multirow{3}{*}{\textbf{SIR}\tnote{a}} 
        & \multicolumn{3}{c}{\textbf{Object}} 
        & \multirow{3}{*}{\textbf{Measurement}} \\
        \cmidrule(lr){5-7}
        & & & & \textbf{Composition} 
        & \begin{tabular}[c]{@{}l@{}} \textbf{BI-RADS} \\ \textbf{Density Distribution}\end{tabular} 
        & \textbf{Lesion Distribution} & \\
        \midrule
        Dataset 1 & & \ding{55} & \ding{55} & & & & \\
        Dataset 2 & 11,024
                  & \checkmark & \ding{55} 
                  & \begin{tabular}[l]{@{}l@{}} Initial pressure map \\ Speed-of-sound map \\ Acoustic density map \\ Acoustic attenuation \\ coefficient map \end{tabular}
                  & \begin{tabular}[l]{@{}l@{}}Type A: 930\\Type B: 2,637\\Type C: 3,243\\Type D: 1,368\end{tabular} 
                  & \begin{tabular}[l]{@{}l@{}}Present: 2,756 \\Paired absent: 2,756 \\Unpaired absent: 5,512\end{tabular} 
                  & \begin{tabular}[l]{@{}l@{}}Sampling rate: 40 MHz\\Geometry: ring array\\Densely sampled, \\ noiseless data\end{tabular} \\
        Dataset 3 &  & \checkmark & \checkmark & & & & \\
        \bottomrule
    \end{tabular}
    }
    \begin{tablenotes}
        \footnotesize
        \item[a] SIR: transducer spatial impulse response.
    \end{tablenotes}
    \end{threeparttable}
\end{table}

Each dataset comprises 11,024 samples, with each sample consisting of an object and its corresponding measurement data. Specifically, the dataset includes 2,756 lesion-present samples, 2,756 paired lesion-absent samples that serve as matched controls for the lesion-present samples in task-based IQ assessment, and 5,512 unpaired lesion-absent samples that provide additional anatomical diversity for training DL models. To reflect clinical breast density variations, the objects span four Breast Imaging Reporting and Data System (BI-RADS) categories \cite{sickles2013acr}: (A) almost entirely fatty, (B) scattered fibroglandular density, (C) heterogeneously dense, and (D) extremely dense breast tissue. A summary of the dataset composition is provided in Table \ref{tab:datasets_overview}. 

The entire dataset generation pipeline consists of three main stages \cite{ChenSPIE2025}: (1) generation of 3D %numerical breast phantoms (
NBPs %) 
and simulation of their initial pressure distributions, (2) generation of 2D objects, %slices, 
and (3) simulation of acoustic measurements using a 2D forward model. 

\subsubsection{3D Numerical Breast Phantoms Generation and Initial Pressure Simulation}
\label{sec:3-1-1}
\begin{figure}[!htbp]
\begin{center}
\begin{tabular}{c} 
\includegraphics[width=0.35\textwidth]{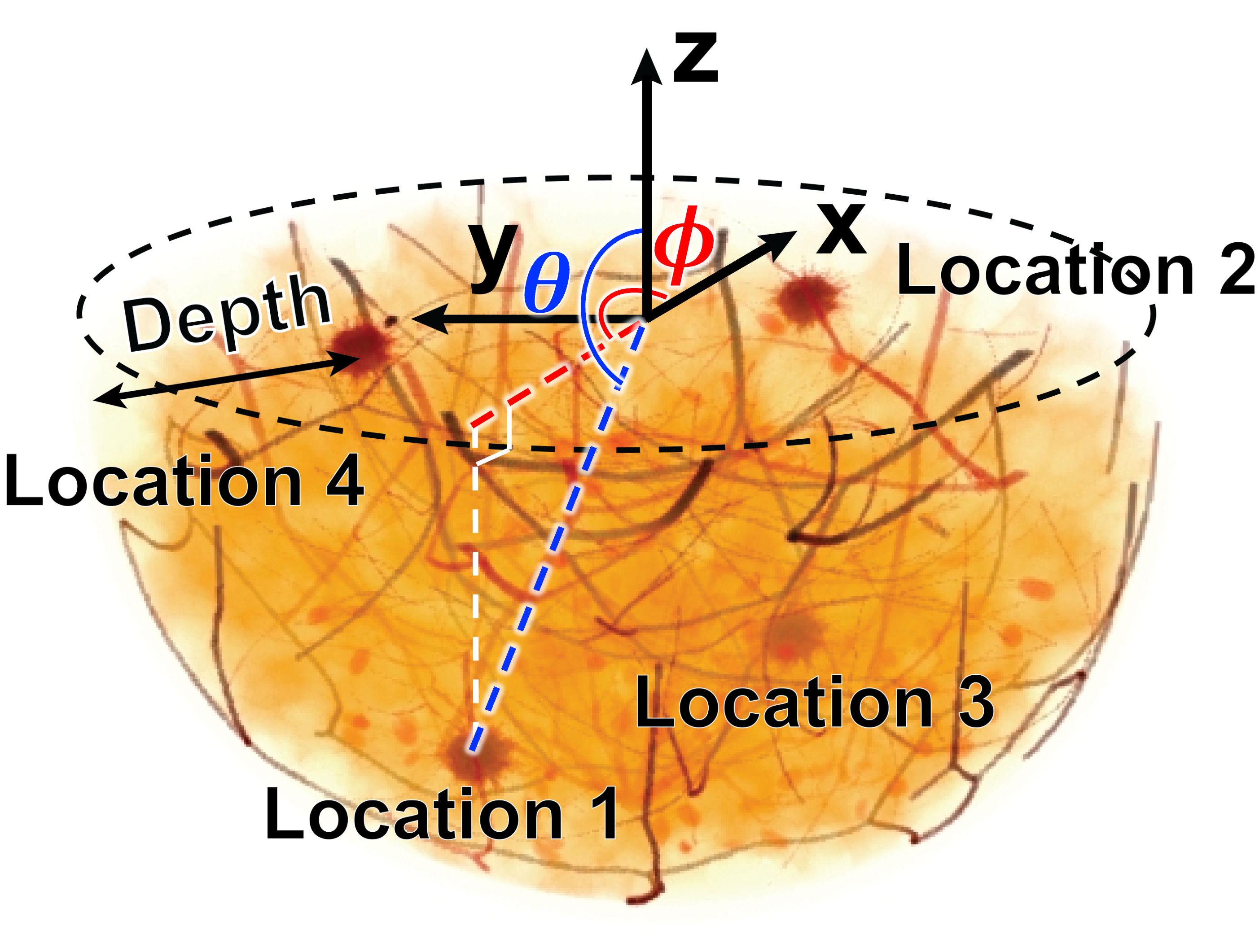}
\end{tabular}
\end{center}
\caption[3D Breast Tissue Label Map with a Numerical Lesion Model inserted in Four Predetermined Locations.] 
{\label{fig:lesion-location} 3D breast tissue label map with a numerical lesion model inserted in four predetermined locations. Lesions at Locations 1 and 2 are positioned at a depth of 10 mm, and those at Locations 3 and 4 at 20 mm, representing shallow and deep cases for assessing light fluence attenuation with depth. Polar angles ($\theta$) of 100$^\circ$  and 140$^\circ$ correspond to regions near the chest wall and nipple, respectively, enabling evaluation under distinct illumination conditions, while azimuthal angles ($\phi$) are separated to minimize optical interference within the same NBP.}
\end{figure}

The stochastic 3D NBPs developed by Park et al. \cite{park2023stochastic} formed the basis of the proposed datasets, which were designed to faithfully represent various breast types through their clinically relevant variability in anatomical structures and functional, optical, and acoustic properties. In this study, 1,500 pairs of stochastic 3D healthy and lesion-inserted NBPs were used to generate the objects in the proposed dataset. %for the proposed dataset objects generation. 
Figures \ref{fig:lesion-location} and %Fig. 
\ref{fig:object generation} (a) present the 3D tissue label map of an example NBP, illustrating its anatomical structure. %illustrate the anatomical structure of the NBP through the 3D tissue label map. 
Each NBP corresponds to %incorporates 
one of the four BI-RADS density categories, with statistically defined breast size and shape parameters, physiologically realistic blood vasculature modeling, and diverse tissue compositions including %such as 
skin, muscle, blood vessels, and fat. For lesion-present NBPs, a numerical lesion phantom (NLP) was constructed, which contains a viable tumor cell region characterized by irregular, spiculated boundaries, as well as a necrotic core and a peripheral angiogenesis region. % containing viable tumor cell regions, necrotic cores, and peripheral angiogenesis regions with irregular spiculated boundaries. 

As illustrated in Fig. \ref{fig:lesion-location}, four candidate locations were predefined within each NBP in spherical coordinates relative to the phantom center: %the NLP was inserted at four predetermined locations within each NBP, defined in spherical coordinates relative to the phantom center: 
Location 1 (10 mm, 140$ ^\circ$, 180$^\circ$), Location 2 (10 mm, 100$^\circ$, 0$^\circ$), Location 3 (20 mm, 140$^\circ$, 270$^\circ$), and  Location 4 (20 mm, 100$^\circ$, 90$^\circ$), where the coordinates represent depth, polar angle $\theta$, and azimuthal angle $\phi$, respectively. For each NBP, the lesion was randomly inserted at one to four of the predefined candidate locations. These locations were selected to systematically investigate depth-dependent optical fluence attenuation at 10 mm and 20 mm, with polar angles chosen to capture distinct illumination conditions: lesions positioned at 100$^\circ$ near the chest wall experience reduced fluence due to limited light penetration, whereas those at 140$^\circ$ closer to the nipple region receive enhanced illumination. The azimuthal angles were selected to %spatially isolate the lesions, ensuring 
ensure that the optical fluence distribution around each lesion would remain independent and minimally affected by neighboring lesions within the same NBP.
%The generation process of 3D NBPs consisted of two main stages, as illustrated in Fig. \ref{fig:object generation} (a). First, the 3D tissue label maps of NBPs were constructed to delineate anatomical structures for subsequent property assignment. Each NBP incorporates one of four BI-RADS density categories with statistically defined breast size and shape parameters, physiologically realistic blood vasculature modeling, and diverse tissue compositions such as skin, muscle, blood vessels, and fat.
% The NLP was inserted at two different depths and azimuthal angles within the breast volume. 

Through stochastic property assignment, 3D distributions of optical and acoustic properties were obtained. %Following the stochastic assignment of optical and acoustic properties, the 3D initial pressure and acoustic property distributions were generated. 
To simulate initial pressure distributions, photon transport simulations were performed in 3D using MCX software \cite{fang2009monte}, based on the 3D optical property distributions and an assumed configuration of the light delivery system. %by mimicking a light delivery PACT system. 
The illumination geometry included 20 evenly positioned arc-shaped illuminators (radius of 145 mm, central angle of 80°), each equipped with five linear fiber-optic segments along its %the arc-shaped 
surface. Each segment was represented by a line beam with a conical angular distribution characterized by a %broadening slit light source (
half-angle of 12.5$^\circ$. %). 
Additional details on property assignment and optical simulation can be found in \citenum{park2023stochastic, park2025virtual}.

\subsubsection{Generation of 2D Objects}% Slices} %Extraction}

\begin{figure}[!htbp]
\begin{center}
\begin{tabular}{c} 
\includegraphics[width=0.98\textwidth]{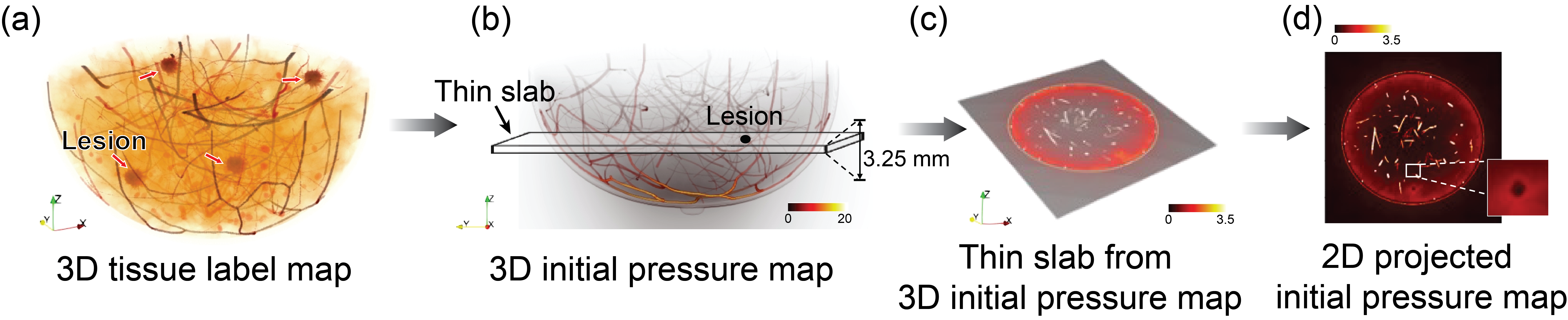}
\end{tabular}
\end{center}
\caption[2D Object Generation Process] 
{\label{fig:object generation} 2D object generation process: (a) A 3D tissue label map of a stochastic NBP is used to assign optical and acoustic properties. (b) A 3D initial pressure map is generated via photon transport simulation based on the optical property distributions. (c) A thin slab is extracted from the 3D initial pressure map. (d) The slab is averaged along the z-axis to produce a 2D initial pressure map.}
\end{figure}
\begin{figure} [h]
\begin{center}
\begin{tabular}{c} 
\includegraphics[width=0.80\textwidth]{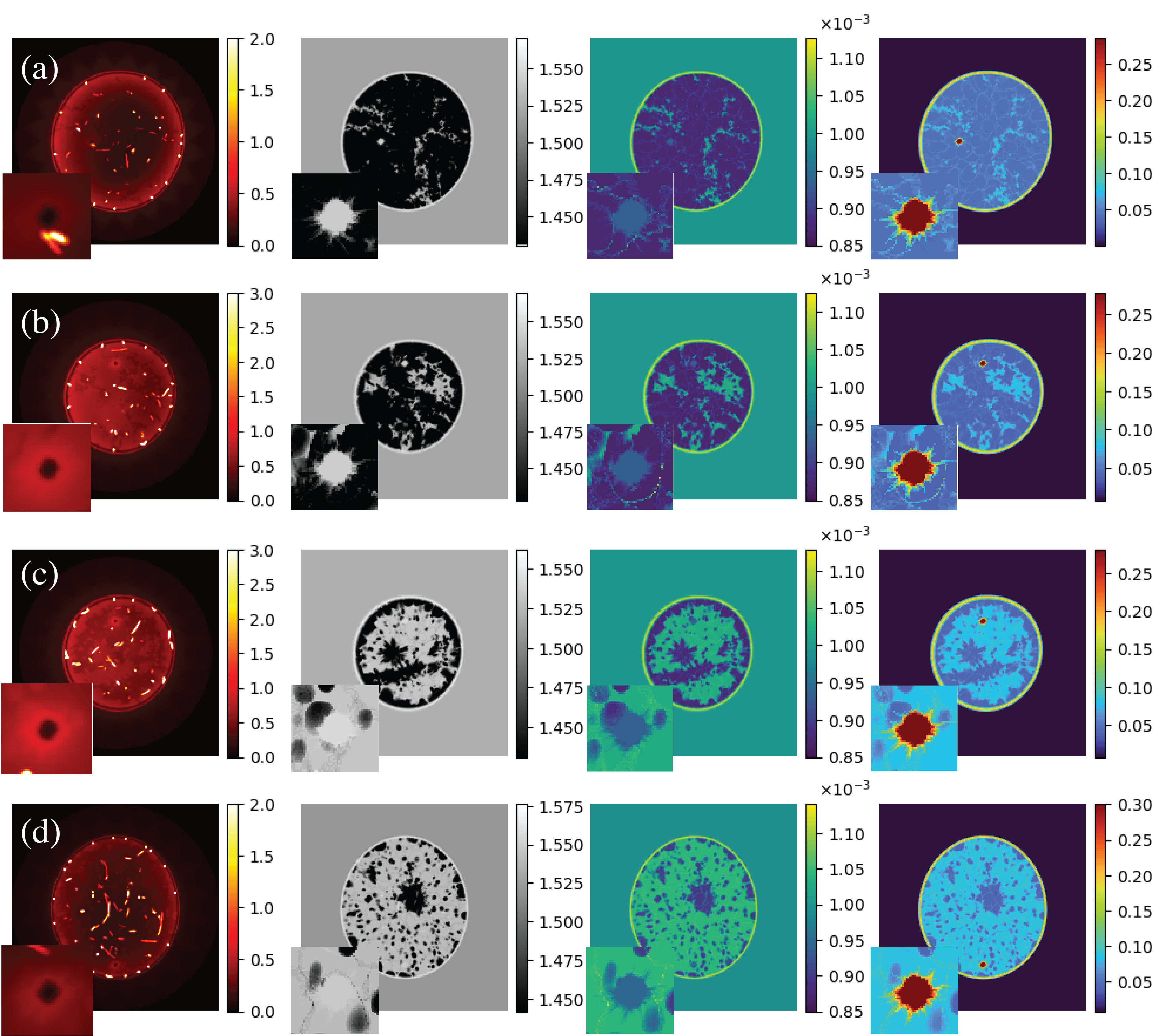}
\end{tabular}
\end{center}
\caption[Representative 2D Lesion-Present Object Examples] 
{\label{fig:example-2d-object} Representative 2D lesion-present object examples from four breast density types. From left to right: initial pressure distribution, SOS, acoustic density map, and attenuation coefficient map. Lesion region is zoomed in. From up to bottom: (a) almost entirely fatty, (b) scattered areas of fibroglandular density, (c) heterogeneously dense, and (d) extremely dense.}
\end{figure}

As shown in Figs. \ref{fig:object generation} (b)-(d), 2D objects were generated from the 3D initial pressure maps and corresponding 3D acoustic property maps to construct the proposed datasets. To approximately emulate the effect of the elevational focusing effect of transducers in 2D PACT systems, 2D initial pressure distributions were derived by averaging thin slabs (3.25 mm thickness) from the 3D initial pressure maps along the z-axis, rather than extracting single slices. Three types of 2D initial pressure distributions were generated based on the $z$-position of the selected 3D slabs relative to the predetermined candidate lesion locations. Slabs centered at these candidate $z$-positions were used to produce \textit{paired} datasets, consisting of \textit{paired lesion-present} and \textit{lesion-absent} slices derived from corresponding 3D lesion-present and healthy phantoms, respectively. These matched pairs enabled direct comparison in signal detection analysis. In contrast, slabs extracted from 3D healthy phantoms and centered at $z$-positions offset by $\pm$5 mm relative to the candidate lesion locations were used to generate \textit{unpaired lesion-absent} slices that  introduced additional anatomical variability into the datasets. 
%whether the selected slab contained a lesion and its position relative to the lesion center: lesion-present slices, paired lesion-absent slices, and unpaired lesion-absent slices. Lesion-present slices were generated from slabs centered at the lesion location and extracted from the 3D lesion-present phantoms. Paired lesion-absent slices were generated from slabs at identical positions but extracted from the corresponding 3D healthy phantoms, creating matched pairs for direct comparison in signal detection analysis. Unpaired lesion-absent slices were generated from slabs extracted from the same 3D healthy phantoms but with slab centers offset by $\pm$5 mm along the z-axis relative to the lesion position, thereby introducing anatomical variability.}
% Based on the z-coordinate of the slab center relative to lesion locations, three types of 2D initial pressure distributions were generated: lesion-present slices, paired lesion-absent slices, and unpaired lesion-absent slices. Here, "\textit{paired}" refers to slices that are generated from the exact same anatomical location (i.e., slab centered at the lesion position) as their corresponding lesion-present slices, but from volumes without lesions, creating matched pairs for direct comparison in signal detection analysis. "\textit{Unpaired}" refers to slices that have no corresponding lesion-present counterpart and are derived from {lesion-absent volumes with slab centers offset by ±5 mm along the z-axis relative to the lesion position, to introduce anatomical variability. 
Similarly, 2D acoustic property maps were obtained by averaging thin slabs from the corresponding 3D acoustic property maps along the z-axis. This approach was motivated by prior work \cite{li20233d}, which demonstrated that %the 
effective 2D acoustic property distributions %map 
can be approximated as %a 
weighted averages of the underlying 3D distributions.
%Similarly, 2D acoustic property maps were generated from the corresponding 3D acoustic property maps. 
Each resulting object was characterized by 2D maps of initial pressure, SOS, density, and acoustic attenuation, all with %at 
a resolution of 1360 $\times$ 1360 pixels and %with 
a pixel size of 0.125 mm. Figure \ref{fig:example-2d-object} shows the representative 2D lesion-present objects %examples from 
corresponding to the four breast density types. 

\subsubsection{%Virtual 
Acoustic Measurements Simulation Using 2D Forward Model}
\begin{figure} [!htbp]
\begin{center}
\begin{tabular}{c} 
\includegraphics[width=0.95\textwidth]{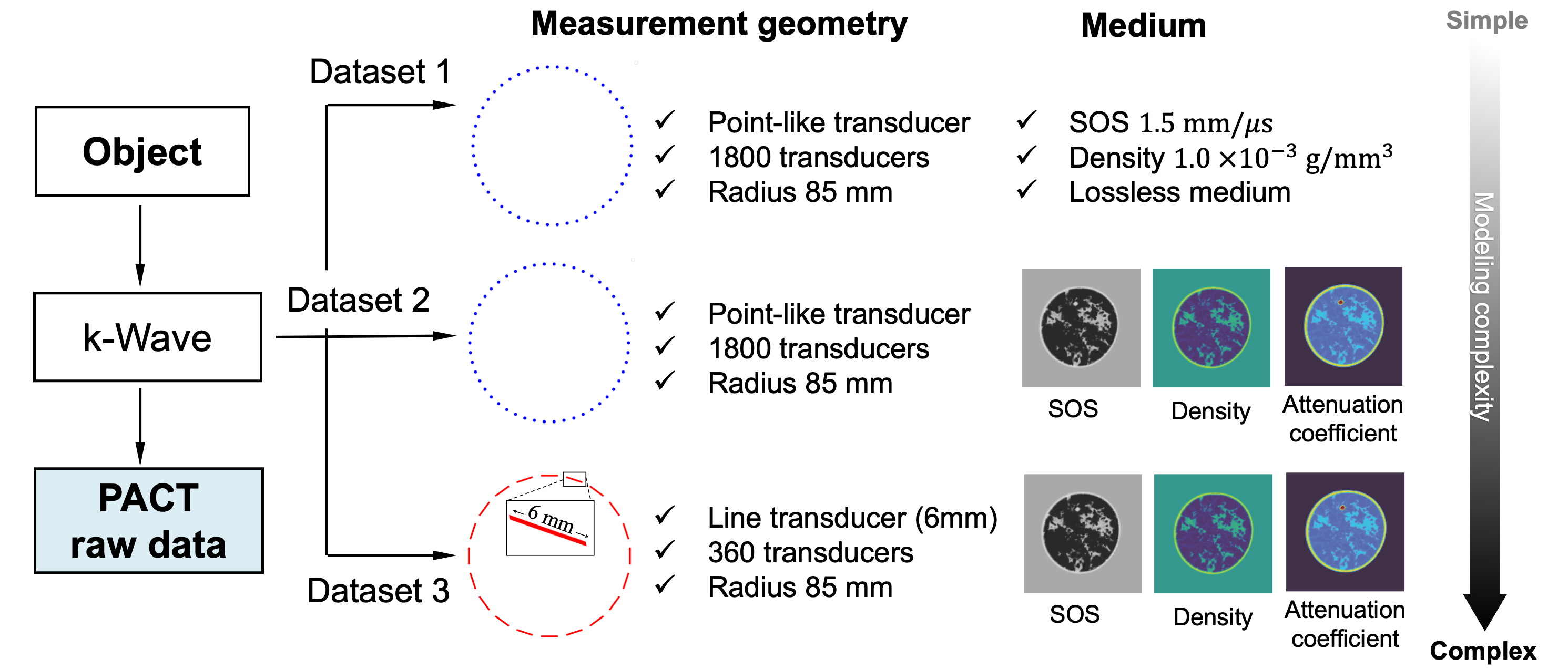}
\end{tabular}
\end{center}
\caption[Virtual PACT Measurement Acquisition for Three Datasets] 
{\label{fig:measurement-generation} Virtual PACT measurement acquisition designs for the three datasets. Three datasets represent increasing modeling complexity, from an acoustically homogeneous medium with point-like transducers to an acoustically heterogeneous medium with finite-size line transducers.}
\end{figure}

As shown in Fig.\ \ref{fig:measurement-generation}, PACT measurement data were acquired for each object via 2D acoustic wave propagation simulation using the k-Wave toolbox \cite{treeby2010k}, assuming a circular measurement geometry with a radius of 85 mm. The sampling rate was set to 40 MHz with a computational grid size of 0.125 mm. Dataset 1 represents a highly stylized scenario with 1,800 point-like transducers uniformly distributed along %around 
the measurement circle and an acoustically homogeneous lossless medium, characterized by a uniform SOS of 1.5 $\mathrm{mm/\mu s}$ and a density of 1$\times 10^{-3}~\mathrm{g/mm^3}$. Dataset 2 introduces greater realism by incorporating spatially varying SOS, density, and acoustic attenuation properties from the extracted 2D acoustic property maps while retaining the 1,800 point-like transducer configuration. Dataset 3 employs the most realistic model, including both acoustic heterogeneity and the SIR of the finite-sized transducers (6 mm in length) arranged in a 360-element ring array. %with a ring array of 360 transducers. 

\subsubsection{Statistical Properties of Generated Breast Datasets}
The datasets incorporate stochastic variations in anatomical structures and acoustic parameters to represent physiological variations in breast tissue. Table \ref{tab:anatomical_acoustical} presents the statistical distributions of tissue parameters across all generated objects. Here, $S$ denotes the breast cross-sectional area, and $f_{fg}$ represents the breast density percentage (i.e., the ratio of fibroglandular tissue to total breast area). 
%, $c$ is the SOS, $\rho$ is the acoustic density, and $\alpha$ is the acoustic attenuation coefficient. 
This variability provides %enables 
a comprehensive representation of breast tissue characteristics to support the training of DL models as well as the development and evaluation of image reconstruction methods.
\begin{table}[htbp]
\centering
\caption{Statistical Distribution of Tissue Parameters in the Generated Breast PACT Datasets}
\resizebox{0.80\linewidth}{!}{%
\begin{tabular}{lccccc}
\toprule
{} & \multicolumn{2}{c}{\textbf{Anatomical Information}} & \multicolumn{3}{c}{\textbf{Acoustical Information}} \\
\cmidrule(r){2-3} \cmidrule(l){4-6}
 & \begin{tabular}{c}S \\ {[mm$^2$]} \end{tabular} 
 & \begin{tabular}{c}$f_{fg}$ \\ {[\%]} \end{tabular} 
 & \begin{tabular}{c}$c_0$ \\ {[mm/$\mu$s]} \end{tabular} 
 & \begin{tabular}{c}$\rho_0$ \\ {[g/mm$^3$]} \end{tabular} 
 & \begin{tabular}{c}$\alpha_0$ \\ {[dB/MHz$^y\cdot$mm]} \end{tabular} \\
\midrule 
Mean & 9298.67  & 58.30 & 1.48 & 9.66$\times 10^{-4}$ & 6.09$\times 10^{-2}$ \\
Std  & 2699.79  & 19.60 & 0.03 & 4.43$\times 10^{-5}$ & 1.03$\times 10^{-2}$ \\
Min  & 3380.97  & 17.40 & 1.43 & 8.47$\times 10^{-4}$ & 3.60$\times 10^{-2}$ \\
Max  & 18596.97 & 89.10 & 1.55 & 1.07$\times 10^{-3}$ & 9.77$\times 10^{-2}$ \\
\bottomrule
\end{tabular}%
}
\label{tab:anatomical_acoustical}
\end{table}
% \subsection{Proposed Evaluation Framework}
% The proposed framework employs both physics-based and task-based IQ assessment metrics to comprehensively evaluate reconstruction algorithm performance. This dual approach captures both traditional image fidelity measures and diagnostic utility, providing a thorough characterization of algorithm capabilities.

% \subsubsection{traditional IQ Assessment}

% \subsubsection{Task-based IQ Assessment}

\subsection{Proposed Evaluation Strategies}
\label{proposed_eval_strateties}
The proposed evaluation strategies include both traditional and task-based IQ assessment metrics to enable clinically meaningful evaluations of reconstruction performance.

\subsubsection{Traditional Measures of Image Quality}
\label{traditional_IQ_measures}
Traditional measures of IQ provide full-reference quantitative assessments of reconstruction fidelity. The study incorporates two standard metrics: relative %Mean Squared Error (
MSE %) 
and %Structural Similarity Index Measure (
SSIM %) 
\cite{wang2004image}. The relative MSE was computed as the MSE divided by the mean value of the true object, accounting for the stochastic variability in breast types and tissue properties within the datasets, and is given by:
\begin{equation}
\text{Relative MSE} = \frac{\|\mathbf{f}^\text{true} - \mathbf{f}\|^2_2}{\|\mathbf{f}^\text{true}\|^2_2} = \frac{\sum_{i=1}^{N}(f^\text{true}_{i} - f_i)^2}{\sum_{i=1}^{N}(f^\text{true}_{i})^2},
\end{equation}
where $\mathbf{f}^\text{true} \in \mathbb{R}^N$ denotes the true object and $\mathbf{f} \in \mathbb{R}^N$ represents the reconstructed image, with $N$ denoting the number of pixels in the discretized image representation. For simplicity, relative MSE is referred to as MSE throughout %in 
this study.

\subsubsection{Task-based Measures of Image Quality}
\label{task-based_IQ_measures}
To assess the diagnostic relevance of the reconstructed images, each reconstruction method was evaluated on two diagnostic tasks using datasets containing objects with lesions positioned at one of %within 
four %possible 
predefined lesion sites (as detailed in Section~\ref{sec:3-1-1}): (1) a binary signal detection task with a known lesion location and (2) a more challenging detection-localization task in which the lesion's specific site among the four predefined candidates was unknown. %with unknown lesion locations. 

\paragraph{Binary Detection Task}
Let $\mathbf{f} \in \mathbb{R}^N$ denote the reconstructed image obtained by a specific reconstruction method, where $N$ is the total number of image voxels. In the binary detection task with a known lesion location,
% For the binary detection task at a known lesion location, 
the objective is to classify the reconstructed image $\mathbf{f}$ as satisfying either a signal-absent hypothesis $H_0$ or the signal-present hypothesis $H_1$, defined as: %. These hypotheses are defined as follows: 
\begin{align}
H_0 &: \mathbf{f} = \mathbf{f}_b + \mathbf{n}&& \text{(lesion absent)}, \\
H_1 &: \mathbf{f} = \mathbf{f}_b + \mathbf{f}_s + \mathbf{n}&& \text{(lesion present)},
\end{align}
where $\mathbf{f}_b \in \mathbb{R}^N$ denotes the background, $ \mathbf{n} \in \mathbb{R}^N$ represents noise, and $\mathbf{f}_s \in \mathbb{R}^N$ represents the signal (lesion). It is assumed that the lesion location is known and that both the background and lesion amplitudes vary statistically with known distributions. %are random and statistically independent. 
This corresponds to a signal-known-statistically and background-known-statistically (SKS/BKS) task with the location-known-exactly (LKE) condition. The BKS formulation is adopted over the simpler background-known-exactly assumption %(BKE) model 
to better reflect realistic object variability. In PACT, signal amplitudes depend not only on the intrinsic optical properties of lesions but also on the local optical fluence, which is determined by photon transport through a stochastically varying background. As a result, even when the lesion location and optical properties are known exactly, the lesion amplitude in the initial pressure distribution varies across samples and cannot be treated as exactly known. By adopting an SKS/BKS task with the LKE condition, the effects of lesion location-dependent optical fluence on lesion detectability can be systematically examined under a specific illumination configuration. %, while the SKS assumption arises from illumination-dependent variations in optical fluence that preclude a fixed lesion amplitude. Moreover, the assumption of known lesion location enables systematic investigation of the effects of illumination pattern and lesion depth on lesion detectability.}
% \textcolor{cyan}{By targeting lesions at specific locations, this formulation is adopted to enable systematic investigation of the effects of illumination pattern and lesion depth on lesion detectability.} 

In this study, a widely used linear observer, the CHO, is employed as the model observer. %\textcolor{cyan}{The CHO can be viewed as an approximation of the optimal linear observer, the HO, that emphasizes features most relevant to the detection task \cite{barrett2013foundations, plativsa2011channelized}, as described in Section \ref{sec:2-3}.} 
It projects $\mathbf{f}$ onto a lower-dimensional channel output vector $\mathbf{v} \in \mathbb{R}^{Q}$ using a channel matrix $\mathbf{T} \in \mathbb{R}^{Q \times N}$, where $Q$ denotes the number of channels and $Q < N$, as follows \cite{plativsa2011channelized}:
\begin{equation}
\label{eq:cho}
\mathbf{v} = \mathbf{T}\mathbf{f}.
\end{equation}
The CHO test statistic is %can 
then %be 
computed as:
\begin{equation}
t(\mathbf{f}) = \mathbf{w}^T \mathbf{v},
\end{equation}
where the CHO template $\mathbf{w} \in \mathbb{R}^Q$ is given by:
\begin{equation}
\label{eq:cho_template}
\mathbf{w} =\left[\frac{1}{2}\left(\mathbf{K}_0+\mathbf{K}_1\right)\right]^{-1} \Delta \overline{\mathbf{v}},
\end{equation}
with %where 
$\mathbf{K}_0$ and $\mathbf{K}_1$ denoting the covariance matrices of $\mathbf{v}$ under hypotheses $H_0$ and $H_1$, respectively. Here, %and 
$\Delta \overline{\mathbf{v}} = \mathbb{E}[\mathbf{v}|H_1] - \mathbb{E}[\mathbf{v}|H_0]$ represents %ing 
the difference between the conditional means of $\mathbf{v}$ under the two hypotheses. The channel matrix $\mathbf{T}$ was defined according to %specified based on 
the selected channel function. 
%\textcolor{red}{The use of channels reduces the high-dimensional image data to a tractable low-dimensional representation while retaining the features most informative for the detection task.} 
Since the objects in the proposed datasets contain radially symmetric lesions, % that exhibit radial symmetry, 
Laguerre-Gauss (LG) channels were adopted in this study \cite{gallas2003validating}. This CHO can be regarded %viewed 
as an approximation of the optimal linear observer, the HO, that emphasizes features most relevant to the detection task \cite{barrett2013foundations, plativsa2011channelized}, as described in Section \ref{sec:2-3}. In practice, the CHO is applied to the regions of interest (ROI) centered on the signal, allowing for adaptive assessment that accounts for anatomical variability \cite{ferrero2017practical}.

With the computed $t(\mathbf{f})$, the observer performance is quantified using the AUC as the FOM. For each lesion location, a distinct CHO template and the corresponding test statistics are computed, and the resulting AUC values are used to quantitatively compare the lesion detection performance of different reconstruction methods. 

\paragraph{Detection-Localization Task}
% The detection-localization task presents greater complexity as the unknown lesion location. 
The detection-localization task introduces greater complexity by requiring identification of an unknown lesion location %s 
among several predefined candidate sites. The observer must classify an image $\mathbf{f}$ as either lesion-absent, %containing no lesion, 
denoted as hypothesis $H_0$, %($H_0$) 
or lesion-present %containing a lesion 
at one of $n_L$ possible locations, denoted as hypothesis $H_{\ell}$, % ($H_{\ell}$, 
where $\ell \in \{1, ..., n_L\}$ with $n_L = 4$ in this study. %). 
The hypotheses are defined as follows:
\begin{align}
H_0 &: \mathbf{f} = \mathbf{f}_b + \mathbf{n}\quad \text{(lesion absent)}, \\
H_{\ell} &: \mathbf{f} = \mathbf{f}_b + \mathbf{f}_{s, \ell} + \mathbf{n}\quad \text{(lesion present at location } \ell \text{)},
\end{align}
where $\mathbf{f}_{s, \ell} \in \mathbb{R}^N$ represents the signal (lesion) at the $\ell$-th candidate location.

A scanning CHO can be adopted for this detection-localization task~\cite{gifford2017visual}. Instead of being applied %Rather than applying the CHO 
to a single ROI as in the binary detection task, the scanning CHO is applied to %extracts 
ROIs extracted from the image $\mathbf{f}$ at all %each of the 
$n_L$ candidate signal locations, computing %and computes 
a location-specific test statistic for each. Specifically, for each candidate location $\ell$, a local test statistic is computed as
\begin{equation}
\label{eq:scanning_cho_t}
t_{\ell}(\mathbf{f}) = \mathbf{w}_{\ell}^T [\mathbf{v} - \mathbf{c}_{\ell}],
\end{equation}
% where $\mathbf{w}_{\ell} = \mathbf{K}_\ell^{-1}\Delta\overline{\mathbf{v}_{\ell}}$ with $\mathbf{K}_\ell = \frac{1}{2}(\mathbf{K}_{0} + \mathbf{K}_{\ell})$, and $\mathbf{c}_{\ell} = \frac{1}{2}(\mathbb{E}[\mathbf{v}|H_0] + \mathbb{E}[\mathbf{v}|H_{\ell}])$ is the location-dependent reference vector that defines the measurement origin~\cite{sen2016accounting}.
where $\mathbf{w}_{\ell} = \mathbf{K}_\ell^{-1}\Delta\overline{\mathbf{v}_{\ell}}$ is the location-specific CHO template. Here, $\Delta\overline{\mathbf{v}_{\ell}} = \mathbb{E}[\mathbf{v}_\ell|H_\ell] - \mathbb{E}[\mathbf{v}_\ell|H_0]$ denotes the difference between the %in 
conditional means of the channel outputs under hypotheses $H_\ell$ and $H_0$, %and 
$\mathbf{K}_\ell = \frac{1}{2}(\mathbf{K}_{0,\ell} + \mathbf{K}_{1,\ell})$ is the average of the covariance matrices of the channel outputs under the two hypotheses, and % at location $\ell$. The term 
$\mathbf{c}_{\ell} = \frac{1}{2}(\mathbb{E}[\mathbf{v}_\ell|H_0] + \mathbb{E}[\mathbf{v}_\ell|H_\ell])$ defines the decision origin, all corresponding to %at 
location $\ell$ ~\cite{sen2016accounting}.
Among all candidate locations, the maximum test statistic is identified as
\begin{equation}
\label{eq:scanning_cho_t_max}
\begin{aligned}
t_{\max }(\mathbf{f})=\max _{\ell} t_{\ell}(\mathbf{f}), \quad r = \arg\max _{\ell} t_{\ell}(\mathbf{f}),
\end{aligned}
\end{equation}
where $r$ denotes the estimated lesion location.

To quantify the detection-localization performance of reconstruction methods, the area under the localization ROC (ALROC) curve is used as the FOM. The localization ROC extends the conventional ROC analysis by incorporating both detection and spatial localization accuracy, plotting the probability of correct localization versus the false positive fraction~\cite{swensson1996unified}.
% The area under the localization ROC (ALROC) curve is used as the FOM to quantify the detection-localization performance of reconstruction methods. 

\section{Benchmarking Study}
\label{benchmarking_study}
The generated datasets of objects and corresponding simulated measurement data, together with the procedures for computing both traditional and task-based measures of IQ, constitute a comprehensive benchmarking framework. This framework, along with all associated data, is publicly available to facilitate reproducible research in PACT reconstruction methods ~\cite{code, data}. To demonstrate the utility of the proposed framework, an example benchmarking study was conducted to evaluate the capability %. This study aims to assess the ability 
of DL-based and physics-based reconstruction methods in compensating for acoustic aberrations in PACT. In this study, it was assumed that both the true objects and the corresponding measurement data were available for determining parameters associated with the DL-based and physics-based methods, while only the measurement data were available for reconstruction during performance evaluation.

\subsection{Dataset Used in Benchmarking Study}
Dataset 2, which incorporates breast acoustic heterogeneity and point-like transducers modeling, was employed in this study. It was systematically partitioned into training, validation, in-distribution (ID) testing, and out-of-distribution (OOD) testing subsets. The ID test set maintained %was designed to match 
the statistical properties of the training distribution, whereas the OOD test set was designed %constructed 
to evaluate model generalization under %to data exhibiting 
distributional shifts. The dataset consists of 4,860 samples, categorized by breast tissue density types and lesion presence. All measurements were simulated using a ring-array geometry with 900 uniformly distributed transducers. Additive Gaussian noise was introduced, modeled as independently and identically distributed (i.i.d.) with a standard deviation equal to 1\% of the maximum photoacoustic signal amplitude across the dataset ~\cite{chen2025learning}.

Table~\ref{tab: dataset_splitting} summarizes the details of the dataset division used for this study. Data corresponding to type D breasts, which represent extremely dense tissue where lesion identification is particularly challenging in gold-standard mammography, were reserved for the OOD test set. The remaining breast types (A, B, and C) were used for training, validation, and ID testing. Specifically, a total of 4,500 samples from breast density types A, B, and C, distributed in a 1:4:4 ratio to reflect clinical prevalence, were allocated such that 70\% were employed for training, 10\% for validation, and 20\% for ID testing. Lesion prevalence was balanced in the training and validation sets but set to 20\% in the ID test set to reflect real-world biopsy-positive rates~\cite{barr2013probably}, resulting in 1,980 lesion-present and 2,520 unpaired lesion-absent samples. A total of 360 type D breast samples were used for OOD testing, matching the total number of type A samples used across the training, validation, and ID test sets. Both the ID and OOD test sets contained 180 lesion-present samples, with the OOD test set assuming balanced lesion prevalence.
\begin{table}[h]
    \centering
    \caption{Dataset Splitting for Benchmarking Study}
    \label{tab: dataset_splitting} 
    \begin{tabular}{lccc}
        \toprule %\hline
        \textbf{Dataset} & \textbf{Size}, $N_\text{samp}$ & \textbf{Breast type} ($N_\text{samp}$) & \textbf{Prevalence} ($N_\text{lp}$, $N_\text{la}$) \\
        \midrule %\hline  \hline
        Training & 3,150 & A (350), B (1400), C (1400) & 50\% (1575, 1575) \\
        Validation & 450 & A (50), B (200), C (200) & 50\% (225, 225) \\
        ID testing & 900 & A (100), B (400), C (400) & 20\% (180$^\text{a}$, 720) \\
        OOD testing & 360 & D (360) & 50\% (180$^\text{a}$, 180) \\
        \bottomrule %\hline
    \end{tabular}
    \caption*{\footnotesize $N_\text{samp}$: Number of samples; $N_\text{lp}$: Number of lesion-present samples; $N_\text{la}$: Number of \textit{unpaired} lesion-absent samples \\ $^\text{a}$An equal number of \textit{paired} lesion-absent samples, corresponding to $N_\text{lp}$ listed in this table, were additionally used for task-based IQ assessments.}
\end{table}

\subsection{Reconstruction Methods}
\label{recon_methods}
The study compared representative physics-based and DL-based reconstruction methods. To avoid inverse crime \cite{Kaipio2007}, a condition in which %where 
the same computational model is used to simulate the measurement data and included as part of the image reconstruction process, the reconstructions were conducted on a coarser grid (0.25 mm grid size, 680 $\times$ 680 pixels) than that employed for the forward simulations (0.125 mm grid size, 1360 $\times$ 1360 pixels).

Two representative DL-based reconstruction methods %used in PACT 
were evaluated, which differ in the domain of operation: data-domain and image-domain learning approaches. %included in this study, distinguished by the domain in which the DL model operates: a data-domain learning approach and an image-domain learning approach.
\begin{itemize}
\item Data-domain learning method ($\mathrm{DL}_{\text {data}}$): Inspired by the U-Net-based sinogram enhancement method \cite{awasthi2020deep}, this approach employs a data-to-data U-Net \cite{ronneberger2015u} to map noisy measurements simulated under an acoustically heterogeneous medium to corresponding noiseless measurements obtained under an acoustically homogeneous lossless medium assumption \cite{treeby2010k}. The output from the U-Net is subsequently used as input to the time reversal (TR) method for image reconstruction under the assumption of a homogeneous medium.
\item Image-domain learning method ($\mathrm{DL}_{\text {image}}$): Inspired by the DeepMB method \cite{dehner2023deep}, this method employs an image-to-image U-Net architecture to map time-delayed %images reconstructed 
reconstructions obtained from noisy measurements to the corresponding initial pressure distribution. The network is trained in a supervised manner using the true initial pressure as the target. Unlike some variants of DeepMB, this method does not incorporate SOS encoding. The network output serves as the final reconstruction.
\end{itemize}Three physics-based methods were used for comparison: TR, penalized least-squares with total variation regularization (PLS-TV), and joint reconstruction (JR) methods. 
\begin{itemize}
\item TR method~\cite{treeby2010k}: This method recovers the initial pressure distribution by backpropagating the measurements in reversed time. The reconstruction was implemented using the k-Wave toolbox, which employs a $k$-space pseudospectral time-domain scheme to solve the system of coupled acoustic equations. A two-region SOS model with distinct values for water and breast tissue is assumed, while acoustic density and attenuation are maintained homogeneous and equal %ivalent 
to water throughout the reconstruction domain. The SOS value for breast tissue was optimized by varying it within the reported range of 1.43 $\mathrm{m/\mu s}$ to 1.57 $\mathrm{m/\mu s}$ in 0.005 $\mathrm{m/\mu s}$ increments \cite{hopp2012breast}.
\item PLS-TV method: This method uses the same two-region SOS model as the TR method and is solved using the fast iterative shrinkage-thresholding algorithm (FISTA) \cite{beck2009fast}. The regularization parameter was selected through parameter sweeping and set to $1 \times 10^{-1}$ based on the minimum MSE between the reconstructed and true initial pressure distributions.
\item JR method~\cite{jeong2025revisiting}: This method jointly estimates the initial pressure and a spatially heterogeneous SOS distribution by solving a constrained optimization problem using support, bound, TV-norm constraints.
%The initial pressure was constrained to be non-negative and set to zero outside the region enclosed by the ring array. For SOS, the domain was divided into breast and background regions. 
%A heterogeneous SOS was estimated only within the breast region, while the background SOS was fixed to known values. 
%The breast-region SOS was bounded by the minimum and maximum SOS observed in the training dataset. 
The TV norms of both the initial pressure and SOS estimates were upper-bounded using thresholds set to the 95~th percentile of the corresponding TV-norm distributions computed from the training dataset.
%Density and AA coefficient were held fixed during JR under the same two-region model, with breast-region values set to their mean values in breast regions observed in training dataset. The AA exponent was also held fixed and set to the mean value observed in the training dataset.
%The initial pressure and SOS were initialized to be zero and to the same two-region breast/background definition, respectively, with the breast-region SOS set to the mean breast SOS observed in the training dataset.
% \item JR method~\cite{jeong2025revisiting}: \textcolor{red}{This method simultaneously estimates both the fully heterogeneous SOS and initial pressure distributions}, with object constraints applied to stabilize the estimation. The density and attenuation were %are 
% modeled using a two-region approach, where breast tissue properties are assigned ensemble-mean values obtained through calibration. 
Additional details are available elsewhere\cite{jeong2025revisiting}.
\end{itemize}

\subsection{Image Quality Evaluation}
Following the evaluation strategies described in Section~\ref{proposed_eval_strateties}, both traditional and task-based measures of IQ were employed to quantitatively assess the performance of the reconstruction methods. For traditional IQ measures (Section~\ref{traditional_IQ_measures}), the MSE and SSIM were computed for the reconstructed images ($680\times 680$ pixels) obtained using the DL-based, TR, and PLS-TV methods for %from 
both the ID and OOD test sets summarized in Table~\ref{tab: dataset_splitting}. These metrics are reported  as quantitative references alongside visual inspection results. % as quantitative references. 
The JR method was excluded from the full test set evaluation due to its high computational cost, which arises from the joint estimation of the SOS and initial pressure. Its evaluation was conducted only on the test samples used in the task-based assessment, which form a representative subset of the full test sets employed for the other reconstruction methods (Section~\ref{recon_methods}). The corresponding MSE and SSIM values for the JR method are reported alongside visual inspection results but are excluded from cross-method statistical comparisons.

% Although the JR method was excluded from cross-method statistical comparisons of MSE and SSIM, its corresponding MSE and SSIM values are reported whenever visual inspection results are presented.

Task-based IQ evaluation was performed to assess the diagnostic efficacy of the reconstructed images in terms of binary lesion detection and localization, as summarized in Table~\ref{tab:IQ_overview}. 
\begin{table}[h]
\centering
\caption{Overview of the Task-based Image Quality Evaluation}
\label{tab:IQ_overview}
\resizebox{0.98\textwidth}{!}{%
    \begin{tabular}{l|c|c}
    \toprule %\hline
    \textbf{Task} &
    \textbf{Binary Detection} &
    \textbf{Detection-Localization} \\
    \midrule %\hline
    ROI size &
    \multicolumn{2}{c}{\(50\times 50\) pixels}  \\
    \hline
    Observer type &
    \begin{tabular}[c]{@{}c@{}}CHO separately trained on \\ID and OOD sets (180 pairs each) \end{tabular} &
    \begin{tabular}[c]{@{}c@{}} Scanning CHO using templates \\ derived from binary detection tasks \end{tabular} \\
    % \begin{tabular}[c]{@{}c@{}}Scanning CHO separately trained on \\ID and OOD sets (180 pairs each) \end{tabular} \\
    \hline
    Data for template estimation &
    \multicolumn{2}{c}{\begin{tabular}[c]{@{}c@{}}45 pairs$^\text{a}$ per lesion location (1 lesion-present + 1 lesion-absent; 4 locations total) \end{tabular}} \\
    \hline
    Data for testing &
    \begin{tabular}[c]{@{}c@{}}Same as data for \\ the template estimation \end{tabular} &
    \begin{tabular}[c]{@{}c@{}}45 group$^\text{b}$ per lesion location \\(1 lesion-present + 3 lesion-absent; 4 locations total)\end{tabular} \\
    \hline
    FOM &
    \begin{tabular}[c]{@{}c@{}}AUC$_\ell$ for each lesion location $\ell$ \end{tabular} &
    ALROC \\
    \bottomrule %\hline
    \end{tabular}
}
    \caption*{\footnotesize $^\text{a}$Each pair consists of one lesion-present sample and one lesion-absent sample for the same lesion location. \\$^\text{b}$Each group consists of one lesion-present sample for that lesion location and three lesion-absent samples from the other three locations.}
\end{table}

\begin{figure} [h!]
\begin{center}
\begin{tabular}{c} 
\includegraphics[width=0.95\textwidth]{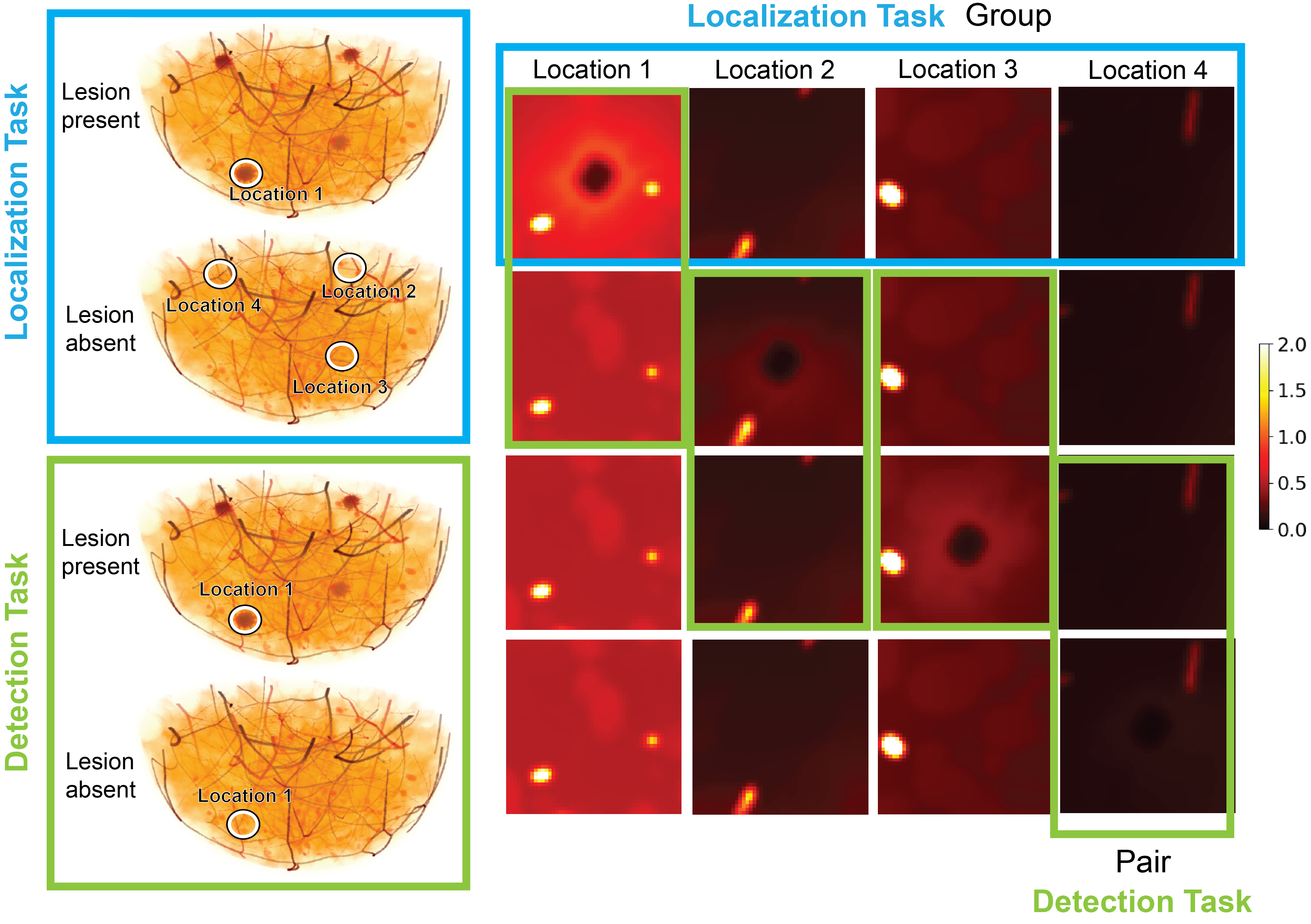}
\end{tabular}
\end{center}
\caption[ROI Image Pairs for the Detection-Localization Task] 
{\label{fig:localization-task-samples} ROI image pairs used for task-based IQ assessment. The left column shows the ROI extraction locations (circles) on 3D numerical breast phantoms for each task. The right panel displays the corresponding ROI images extracted from the true object. For the localization task, each group (cyan boxes) consists of one lesion-present ROI image at a specific location and three lesion-absent images from the other locations. For the detection task, each pair (green boxes) consists of a lesion-present ROI image and a lesion-absent ROI image from the same anatomical location.}
% For the binary lesion detection task, each ROI pair (green boxes) consists of a lesion-present image and a lesion-absent image from the same anatomical location. For the lesion localization task, each sample (blue boxes) includes one lesion-present image at a specific location and three lesion-absent images from the other locations.}}
\end{figure}

\paragraph{Binary Detection Task} 
%The binary detection task was performed separately for each of the four known lesion locations and for both the ID and OOD test sets. At each location, 
ROIs were extracted from the reconstructed images at each of the four lesion locations, and the CHO analysis described in Section~\ref{task-based_IQ_measures} was independently performed for each lesion location %applied 
to quantify detection performance separately on the ID and OOD test sets.
% For each ID and OOD test set and for each of the four exactly known lesion locations, a binary detection task was defined. The performance of the reconstruction methods presented in Section~\ref{recon_methods} was quantified using the CHO analysis described in Section~\ref{task-based_IQ_measures}. 
Specifically, for each lesion location, 45 pairs of lesion-present and lesion-absent ROIs centered on the corresponding lesion location were extracted from reconstructed images in both the ID and OOD test sets. The ROI covered a $50 \times 50$ pixel area and was represented as a vector $\mathbf{f}\in\mathbb{R}^{2500}$ under hypotheses $H_0$ (lesion-absent) and $H_1$ (lesion-present), as defined in Section~\ref{task-based_IQ_measures}. Example ROI pairs for the four lesion locations, derived from the true initial pressure distributions, are indicated with green boxes in Fig.~\ref{fig:localization-task-samples}.
The CHO was configured with four LG channels ($Q=4$) and a Gaussian width of 15 pixels. The CHO test statistics $t(\mathbf{f})$ were computed using observer templates $\mathbf{w}$ derived from Equations~\eqref{eq:cho}--\eqref{eq:cho_template}. Template estimation followed a resubstitution approach \cite{barrett2013foundations}, in which the same ROI pairs used to compute the test statistics were also used for the template estimation. The detection performance was subsequently evaluated through ROC analysis using the RJafroc R package \cite{chakraborty2023rjafroc} to compute AUC based on the test statistics $t(\mathbf{f})$ for each reconstruction method.

\paragraph{Detection-Localization Task} %For each ID and OOD test set, a detection-localization task was defined to identify an unknown lesion location among four predefined candidate sites, $\ell \in \{1, 2, 3, 4\}$. 
As in the binary detection task, ROIs were extracted from the reconstructed images at the four predefined candidate sites, $\ell \in \{1, 2, 3, 4\}$. 
%The performance of the reconstruction methods presented in Section~\ref{recon_methods} on this task was quantified using t
The scanning CHO analysis described in Section~\ref{task-based_IQ_measures} was independently performed for each lesion site to quantify localization performance, specifically identifying the unknown lesion site among the four predefined candidate locations. Unlike the ROI pairs used in the binary detection task, 45 ROI groups for this task were organized by lesion location, with each group consisting of one lesion-present ROI and three lesion-absent ROIs corresponding to the remaining candidate sites, all derived from the same breast. An example group is indicated by the blue box in Fig.~\ref{fig:localization-task-samples}.
%To implement the scanning CHO, ROIs were extracted at each candidate location to compute location-specific test statistics.
The LG channels for the scanning CHO were configured identically to those used in the binary detection task. The location-specific templates $\mathbf{w}_{\ell}$ in Equation~\eqref{eq:scanning_cho_t} were estimated using the ROI pairs employed in the binary detection task. %\textcolor{cyan}{For each of the ID and OOD test sets, 45 ROI groups were used for the localization testing.}
% For testing, 45 ROI pairs were established separately for ID and OOD test sets. 
%Each \textcolor{cyan}{group} contained one lesion-present ROI corresponding to one of the predefined lesion locations and three lesion-absent ROIs corresponding to the remaining locations, as illustrated in Fig.~\ref{fig:localization-task-samples}, where an example \textcolor{cyan}{group} is indicated with a blue box. 
The lesion location $r$ was estimated using $\mathbf{w}_{\ell}$ as formulated in Equations~\eqref{eq:scanning_cho_t} and \eqref{eq:scanning_cho_t_max}, and the task performance metric ALROC was computed using the RJafroc R package~\cite{chakraborty2023rjafroc} for each reconstruction method.

\section{Benchmarking Study Result}
\label{benchmarking_study_result}
\subsection{In-distribution Test Result} 
\begin{figure*} [h!]
\begin{center}
\begin{tabular}{c} 
\includegraphics[width=0.98\textwidth]{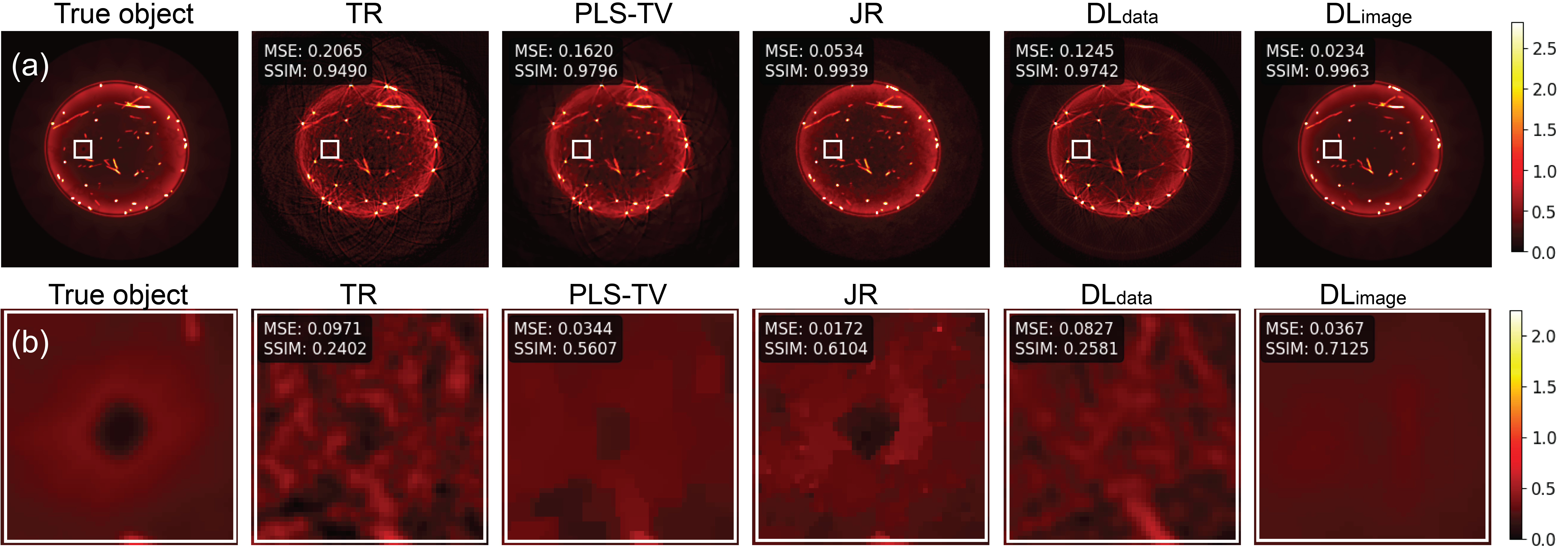}
\end{tabular}
\end{center}
\caption[Visual Assessment of Reconstruction Results for a Representative Sample] 
{\label{fig:visulization} Visual assessment of reconstruction results for a representative sample on the ID test set. The first row (a) presents reconstructed images with annotated MSE and SSIM values, with white boxes marking lesion locations. The second row (b) displays zoomed-in lesion regions. The images, from left to right, show the true initial pressure, and reconstructions from TR, PLS-TV, JR, and $\mathrm{DL}_{\text {data}}$, $\mathrm{DL}_{\text {image}}$. The results demonstrate that, the commonly used IQ metrics yielded favorable scores even when the DL-based reconstructions especially $\mathrm{DL}_{\text {image}}$ failed to recover the lesion, indicating their limited suitability for clinically relevant IQ assessments.}
\end{figure*}

Figures \ref{fig:visulization} (a) and (b) present representative images reconstructed using the %various 
methods considered in this study on the ID test set. Figure \ref{fig:visulization} (a) displays the physical measures of IQ, MSE and SSIM, and it reveals that the DL-based method ($\mathrm{DL}_{\text {image}}$) achieved the highest overall IQ, consistent with the visually %impression of 
reduced distortion and sharper anatomical details compared with the other methods. However, as illustrated in the zoomed-in lesion regions in Fig. \ref{fig:visulization} (b), the $\mathrm{DL}_{\text {image}}$ method failed to recover %reconstruct 
clinically significant lesions that were %are 
successfully reconstructed %identified 
by the physics-based methods %such as 
(TR, PLS-TV, and JR). This discrepancy highlights the limitation of traditional measures of IQ, which may not fully capture diagnostic utility.

Table \ref{tab:ID_MSE_SSIM} reports the MSE and SSIM values for different reconstruction methods on %over %across 
the entire ID test set. Consistent with Fig. \ref{fig:visulization}, the DL-based methods exhibited superior overall IQ. %demonstrate superior performance in overall global IQ.
\begin{table}[h!]
\caption{Performance Evaluation of Different Reconstruction Methods}
\centering
\resizebox{0.60\textwidth}{!}{
\begin{tabular}{p{0.22\columnwidth} c c}
\toprule %\hline
\textbf{Method} & \textbf{MSE} & \textbf{SSIM} \\
\midrule %\hline  \hline
TR          & $0.327 \pm 0.098$  & $0.934 \pm 0.016$ \\
PLS-TV      & $0.267 \pm 0.079$  & $0.975 \pm 0.007$ \\ 
$\mathrm{DL_{data}}$ &  $0.091 \pm 0.040$ & $0.974 \pm 0.010$  \\
$\mathrm{DL_{image}}$ & $0.017 \pm 0.009$ & $0.997 \pm 0.002$ \\
\bottomrule \hline
\end{tabular}}
\label{tab:ID_MSE_SSIM}
\end{table}

\begin{figure*} [h!]
\begin{center}
\begin{tabular}{c} 
\includegraphics[width=0.98\textwidth]{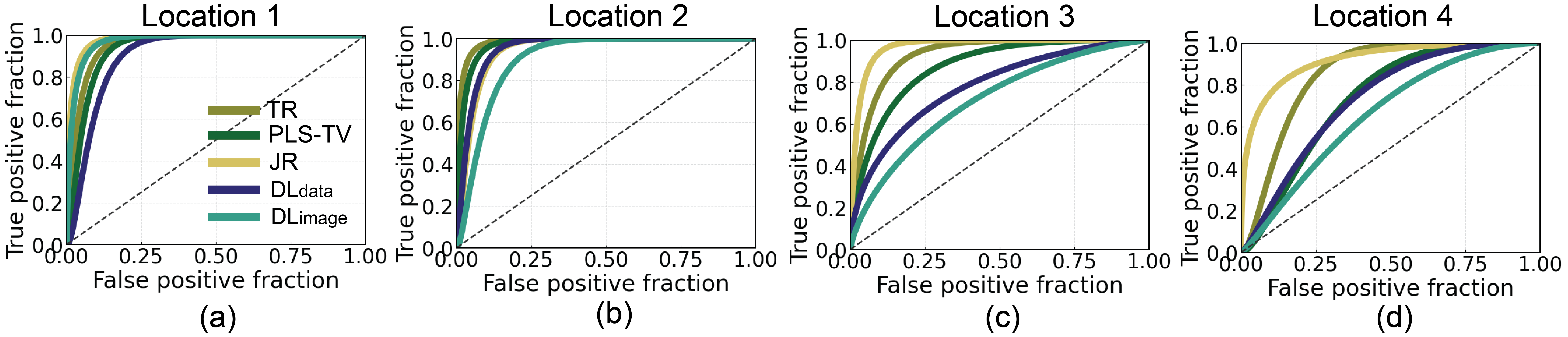}
\end{tabular}
\end{center}
\caption[ROC Analysis of Lesion Detection Performance] 
{\label{fig:ROC-binary-detection} ROC analysis of binary lesion detection performance for reconstruction methods on the ID test set across four lesion locations: (a)–(d) correspond to Locations 1 to 4. The results demonstrate performance degradation from shallow (Locations 1 and 2, 10 mm) to deep (Locations 3 and 4, 20 mm) lesions, where DL-based methods exhibited greater sensitivity to depth-dependent optical attenuation.}
\end{figure*}

\begin{figure*} [h!]
\begin{center}
\begin{tabular}{c} 
\includegraphics[width=0.98\textwidth]{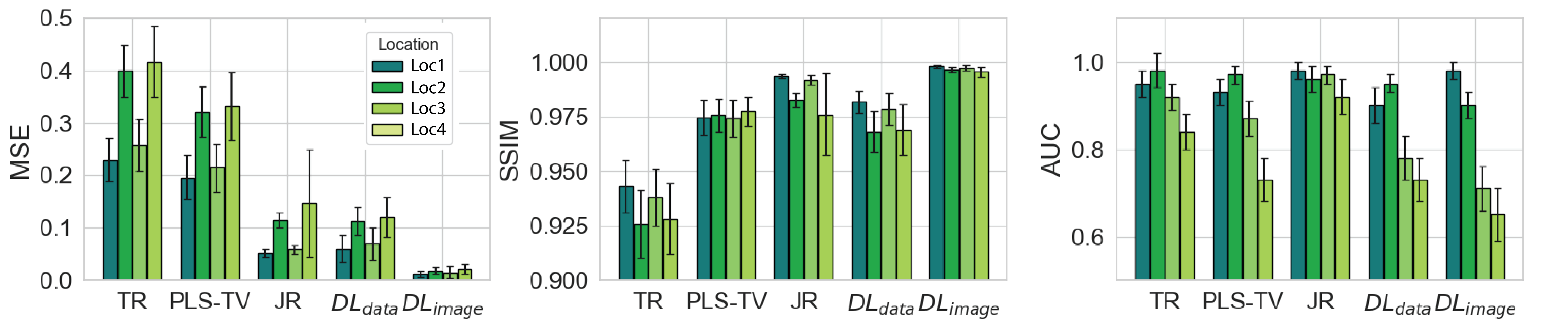}
\end{tabular}
\end{center}
\caption[Bar plots Showing Traditional IQ Metrics and Task-based IQ Metric for Binary Detection Tasks] 
{\label{fig:ID-binary-result} Bar plots showing traditional measures of IQ (MSE and SSIM) and a task-based measure of IQ (AUC) for binary detection tasks at four lesion locations in the ID test set. The results indicate that the traditional metrics failed to capture depth-dependent variations in lesion detectability across methods, whereas AUC clearly revealed performance degradation at greater depths (i.e., Locations 3 and 4), particularly highlighting the reduced lesion detectability of DL-based methods when clinical important low-contrast lesions were missed.}
\end{figure*}

To comprehensively evaluate diagnostic performance, Figs. \ref{fig:ROC-binary-detection} and \ref{fig:ID-binary-result} present the results of binary lesion detection analyses on the ID test set. Figure \ref{fig:ROC-binary-detection} shows the ROC plots for location-specific lesion detection with different reconstruction methods, evaluated using test statistics $t_\ell(\mathbf{f})$ derived from the CHOs, while Fig. \ref{fig:ID-binary-result} reports the corresponding AUC values. For comparison, the MSE and SSIM values for the same ID test samples are also included in Fig. \ref{fig:ID-binary-result}. The results in Fig. \ref{fig:ID-binary-result} show that traditional metrics (MSE and SSIM) consistently favored the DL-based methods across all lesion locations, %show
exhibiting minimal depth-dependent variation. In contrast, the task-based measure of IQ (AUC) revealed significant depth-dependent performance differences across reconstruction methods. Specifically, at a depth of 10 mm %shallow depths 
(Locations 1 and 2), %10 mm), 
AUC values aligned with traditional metrics, indicating superior lesion detectability for DL-based methods. However, at a depth of 20 mm %greater depths 
(Locations 3 and 4), %20 mm), 
MSE and SSIM values did not indicate degraded performance relative to 10 mm, whereas AUC values decreased across all methods, consistent with reduced signal-to-noise ratio due to optical attenuation at greater depths. Notably, the DL-based methods maintained the best MSE and SSIM yet showed the lowest AUC at lesion depth of 20 mm, reflecting their %its 
reduced lesion detectability of low-contrast lesions. In contrast, the physics-based methods achieved higher lesion detectability despite inferior MSE and SSIM, with the JR method demonstrating the best task-based performance at both depths.

\begin{figure} [h!]
\begin{center}
\begin{tabular}{c} 
\includegraphics[width=0.98\textwidth]{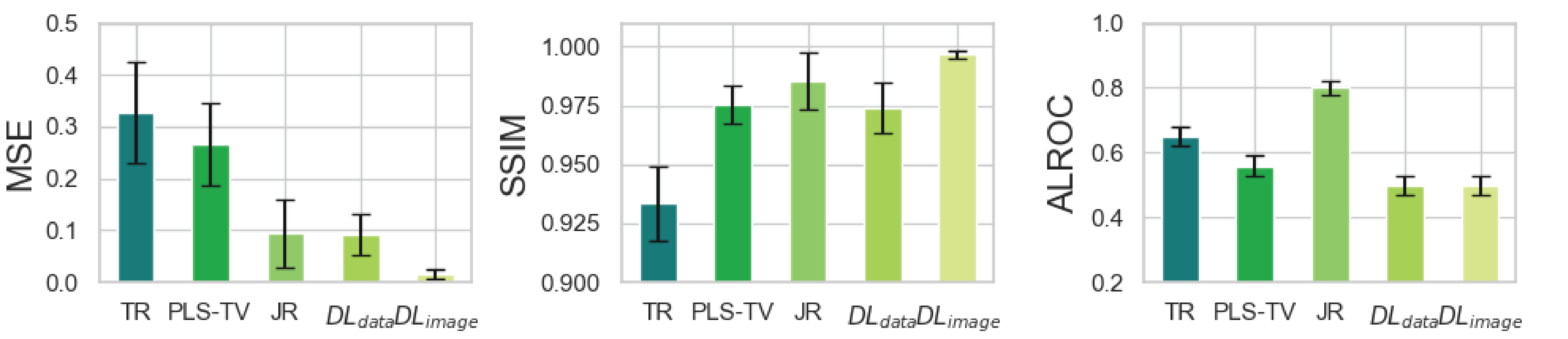}
\end{tabular}
\end{center}
\caption[Bar Plots of Task-based Measure of IQ for Lesion Localization] 
{\label{fig:ID-localization-test} Bar plots of task-based measure of IQ for lesion localization (ALROC) in the ID test set, with traditional measures of IQ (MSE and SSIM) included for comparison. Results show that the DL-based method achieved the highest traditional metrics values but the lowest ALROC values, indicating the poorest localization performance and highlighting the limitations of traditional metrics in assessing lesion localization.}
\end{figure}

Figure \ref{fig:ID-localization-test} presents the results of the lesion localization task, along with the MSE and SSIM values for the ID test set. Consistent with the binary detection results, the DL-based methods yielded the lowest ALROC values despite achieving the highest traditional measures of IQ scores, indicating the poorest localization performance. This deficiency was primarily due to missed lesions at deeper locations, as evidenced in Fig. \ref{fig:ID-binary-result}. In contrast, the JR method achieved significantly superior localization performance despite suboptimal MSE and SSIM scores. These findings emphasize the critical importance of evaluating reconstruction methods using both traditional IQ and diagnostic task performance metrics within the proposed framework. It is particularly crucial for DL-based methods that may achieve superior physical measures of IQ while failing to reconstruct clinically significant low-contrast lesions.

\subsection{Out-of-distribution Test Result} 
\begin{figure} [h!]
\begin{center}
\begin{tabular}{c} 
\includegraphics[width=0.85\textwidth]{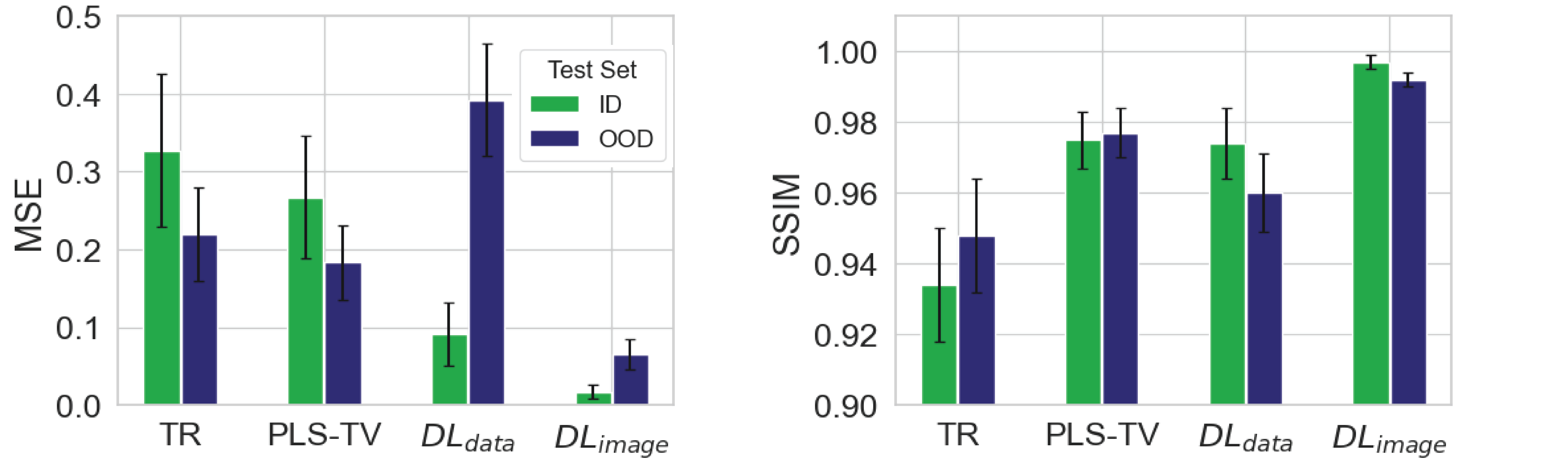}
\end{tabular}
\end{center}
\caption[Bar Plots comparing MSE and SSIM across Reconstruction Methods] 
{\label{fig:ID-OOD-MSE-SSIM} Bar plots comparing MSE (left) and SSIM (right) across reconstruction methods. The quantitative results indicate that the DL-based methods struggled with generalization on the OOD test set, despite dense breasts being generally considered easier to reconstruct, as indicated by the performance trends in the physics-based methods.}
\end{figure}

Figure \ref{fig:ID-OOD-MSE-SSIM} presents bar plots of the MSE and SSIM values for the different reconstruction methods on the OOD test set, with the ID test set included for comparison. The OOD test set consisted exclusively of extremely dense breasts, which are typically considered easier to accurately reconstruct, as evidenced by the improved performance of the physics-based methods from the ID to OOD test results. However, the DL-based methods struggled to generalize to this unseen data distribution, as indicated by degraded traditional measures of IQ metrics. These results suggest that the DL-based approaches exhibited sensitivity to distributional shifts and further highlight the need for meaningful robustness evaluation of DL-based reconstruction methods, particularly regarding their generalizability.

Table \ref{tab:OOD_all_metrics} reports the quantitative results for traditional and task-based measures of IQ on the OOD test set, where bold values indicate the best result per row. In the binary lesion detection task, consistent with the ID test results, traditional metrics suggest that the $\mathrm{DL}_{\text{image}}$ method performed well, particularly at Locations 2 and 4. However, its AUC values indicate limited detectability, especially for lesions located at greater depths (Locations 3 and 4, 20 mm). In contrast, the JR method consistently achieved superior location-specific lesion detectability. The results of the lesion localization follow the same trend: the $\mathrm{DL}_{\text{image}}$ method yielded the lowest ALROC values, despite the competitive MSE and SSIM scores, indicating that the reconstructed images were less informative for the localization task. Conversely, the physics-based methods, although yielding %despite 
lower scores on traditional IQ metrics, significantly outperformed $\mathrm{DL}_{\text{image}}$ in lesion localization accuracy.
\begin{table}[htbp]
\centering
\caption{Quantitative Comparison of Reconstruction Methods (Mean $\pm$ Std) on the OOD Test Set}
\resizebox{0.90\textwidth}{!}{
\begin{tabular}{l c ccccc}
\toprule
\textbf{Metric} & \textbf{Location}
& \textbf{TR} & \textbf{PLS-TV} & \textbf{JR} & \textbf{\(\mathrm{DL_{data}}\)} & \textbf{\(\mathrm{DL_{image}}\)} \\
\midrule\midrule
\multirow{5}{*}{MSE ~$\downarrow$}
& Location 1  & 0.159\std{0.022} & 0.136\std{0.023} & \textbf{0.050\std{0.009}} & 0.354\std{0.084} & 0.061\std{0.016} \\
& Location 2  & 0.265\std{0.046} & 0.219\std{0.037} & 0.080\std{0.014} & 0.436\std{0.055} & \textbf{0.064\std{0.022}} \\
& Location 3  & 0.176\std{0.026} & 0.149\std{0.025} & \textbf{0.055\std{0.008}} & 0.389\std{0.074} & 0.065\std{0.016} \\
& Location 4  & 0.273\std{0.018} & 0.223\std{0.017} & 0.082\std{0.012} & 0.390\std{0.045} & \textbf{0.070\std{0.022}} \\
\addlinespace
\midrule
\multirow{5}{*}{SSIM ~$\uparrow$}
& Location 1  & 0.964\std{0.007} & 0.981\std{0.005} & \textbf{0.996\std{0.001}} & 0.989\std{0.002} & 0.973\std{0.005} \\
& Location 2  & 0.940\std{0.010} & 0.975\std{0.006} & 0.988\std{0.002} & 0.952\std{0.007} & \textbf{0.991\std{0.002}} \\
& Location 3  & 0.954\std{0.010} & 0.978\std{0.006} & \textbf{0.994\std{0.001}} & 0.964\std{0.006} & 0.993\std{0.002} \\
& Location 4  & 0.935\std{0.013} & 0.973\std{0.008} & 0.986\std{0.003} & 0.952\std{0.010} & \textbf{0.991\std{0.003}} \\
\addlinespace
\midrule
\multirow{4}{*}{AUC ~$\uparrow$}
& Location 1  & 0.97\std{0.02} & 0.97\std{0.02} & \textbf{0.98\std{0.01}} & 0.96\std{0.03} & 0.95\std{0.02} \\
& Location 2  & 0.96\std{0.02} & 0.93\std{0.03} & \textbf{0.96\std{0.02}} & 0.92\std{0.03} & 0.85\std{0.04} \\
& Location 3  & 0.90\std{0.04} & 0.90\std{0.04} & \textbf{0.90\std{0.03}} & 0.72\std{0.06} & 0.72\std{0.05} \\
& Location 4  & 0.75\std{0.05} & 0.71\std{0.05} & \textbf{0.78\std{0.05}} & 0.67\std{0.08} & 0.65\std{0.06} \\
\addlinespace
\midrule
ALROC ~$\uparrow$
& Total & 0.67\std{0.03} & 0.60\std{0.03} & \textbf{0.68\std{0.03}} & 0.56\std{0.03} & 0.44\std{0.03} \\
\bottomrule
\end{tabular}
}
\label{tab:OOD_all_metrics}
\end{table}

\section{Discussion and Conclusion}
\label{conclusion}
To facilitate fair comparison and clinical utility assessment of DL-based methods for PACT, this study proposes a benchmarking pipeline comprising realistic synthetic datasets and %evaluation 
strategies to rigorously assess the performance of DL-based acoustic inversion methods. Specifically, the datasets feature breast tissue objects derived from 3D stochastic numerical phantoms that capture anatomical and physiological variability with clinically relevant lesions, paired with measurements generated through high-fidelity virtual imaging that closely mimics real-world conditions. The evaluation strategies combine traditional measures of IQ for quantifying overall image fidelity %quality 
with task-based measures of IQ for assessing lesion detection and localization diagnostic performance. 

The proposed datasets serve as a well-characterized benchmark that offers a flexible and cost-effective alternative to experimental data. Because synthetic datasets are designed with known ground truth, they enable large-scale data generation at relatively low cost through computational modeling. Previous studies have employed synthetic data based on segmented anatomical maps \cite{staal2004ridge}, vascular simulations \cite{hamarneh2010vascusynth}, or their combinations \cite{bench2020toward}. However, to ensure clinical relevance, synthetic objects must accurately reproduce \textit{in vivo} anatomical, physiological, and pathological variations. The stochastic object models in this study incorporate comprehensive variability and support robust feature learning by DL-based models, thereby enhancing representational power and generalizability. Furthermore, high-fidelity virtual imaging is critical for assessing DL-based methods under clinically realistic conditions, allowing repeated evaluation \textit{in silico} before translation to real-world applications. The varying levels of acoustic modeling complexity in the datasets permit systematic investigation of diverse PACT acoustic inversion problems while controlling for other confounding factors. For example, Dataset 1, which excludes acoustic heterogeneity and SIR effects, can be used to study sparse-view PACT reconstruction.

The benchmarking results revealed critical limitations of traditional measures of IQ for clinical assessment and highlighted the importance of incorporating task-based measures of IQ. While Table \ref{tab:ID_MSE_SSIM} showed that the DL-based methods achieved superior MSE and SSIM scores, Fig. \ref{fig:visulization} revealed their limited sensitivity to subtle yet clinically significant lesions, in contrast to physics-based methods that exhibited the opposite trend. This discrepancy demonstrates that traditional measures of IQ relying on pixel-level comparisons may not fully capture diagnostically relevant features, underscoring the value of task-based measures of IQ, which more reliably reflects diagnostic performance. These task-based measures of IQ also provide quantitative guidance for optimizing DL-based methods for clinical deployment. For instance, Figs. \ref{fig:ROC-binary-detection} and \ref{fig:ID-binary-result} illustrated the impact of lesion depth on detectability, revealing that DL methods exhibited particularly poor performance for low-contrast lesions at greater depths. These insights suggest potential improvements, such as %including 
incorporating region-of-interest weighting during training or developing task-specific architectures.

Future work will incorporate additional modeling complexities to further enhance dataset realism and expand the inclusion of %incorporate more 
physical and task-based measures of IQ for more comprehensive %meaningful 
evaluation strategies. A fully 3D extension of the proposed framework will also be developed to accommodate more realistic clinical imaging scenarios. The proposed datasets and evaluation strategies are %will be made 
publicly available through \citenum{data, code} to support benchmarking efforts and facilitate algorithm development within the photoacoustic imaging community, ultimately advancing the clinical translation of PACT technologies.

\section{Acknowledgments}
This work was supported in part by the National Institutes of Health, United States grants EB031772, EB034249, EB031585 and EB034261. This work used the Delta system at the National Center for Supercomputing Applications through allocation MDE230007 from the Advanced Cyberinfrastructure Coordination Ecosystem: Services \& Support (ACCESS) program, which is supported by U.S. National Science Foundation grants \#2138259, \#2138286, \#2138307, \#2137603, and \#2138296. 
The authors would like to thank Hsuan-Kai Huang for the valuable discussions and insightful comments that greatly contributed to this work. The authors acknowledge the use of ChatGPT (OpenAI) for assistance with language editing and grammar refinement.

% \disclosures 
\section*{Disclosures}
The authors declare that there are no financial interests, commercial affiliations, or other potential conflicts of interest that could have influenced the objectivity of this research or the writing of this paper.

\section* {Code and Data Availability} 
The developed synthetic breast datasets and scripts for the evaluation strategies will be made publicly available through open-access repositories \citenum{data, code} upon acceptance of the paper.

%%%%% References %%%%%
\bibliography{reference}   

@inproceedings{chen2025benchmarking,
  title={Benchmarking deep learning-based reconstruction in photoacoustic computed tomography with clinically relevant synthetic datasets},
  author={Chen, Panpan and Park, Seonyeong and Jeong, Gangwon and Cam, Refik Mert and Huang, Hsuan-Kai and Villa, Umberto and Anastasio, Mark A},
  booktitle={Photons Plus Ultrasound: Imaging and Sensing 2025},
  volume={13319},
  pages={70--76},
  year={2025},
  organization={SPIE}
}

@article{awasthi2020deep,
  title={Deep neural network-based sinogram super-resolution and bandwidth enhancement for limited-data photoacoustic tomography},
  author={Awasthi, Navchetan and Jain, Gaurav and Kalva, Sandeep Kumar and Pramanik, Manojit and Yalavarthy, Phaneendra K},
  journal={IEEE transactions on ultrasonics, ferroelectrics, and frequency control},
  volume={67},
  number={12},
  pages={2660--2673},
  year={2020},
  publisher={IEEE}
}

@article{swensson1996unified,
  title={Unified measurement of observer performance in detecting and localizing target objects on images},
  author={Swensson, Richard G},
  journal={Medical physics},
  volume={23},
  number={10},
  pages={1709--1725},
  year={1996},
  publisher={Wiley Online Library}
}

@inproceedings{ferrero2017practical,
  title={Practical implementation of channelized hotelling observers: effect of ROI size},
  author={Ferrero, Andrea and Favazza, Christopher P and Yu, Lifeng and Leng, Shuai and McCollough, Cynthia H},
  booktitle={Proceedings of SPIE--the International Society for Optical Engineering},
  volume={10132},
  pages={101320G},
  year={2017}
}

@inproceedings{li20233d,
  title={3D full-waveform inversion in ultrasound computed tomography employing a ring-array},
  author={Li, Fu and Villa, Umberto and Duric, Nebojsa and Anastasio, Mark A},
  booktitle={Medical Imaging 2023: Ultrasonic Imaging and Tomography},
  volume={12470},
  pages={99--104},
  year={2023},
  organization={SPIE}
}

@article{wang2023photoacoustic,
  title={Photoacoustic imaging with limited sampling: a review of machine learning approaches},
  author={Wang, Ruofan and Zhu, Jing and Xia, Jun and Yao, Junjie and Shi, Junhui and Li, Chiye},
  journal={Biomedical optics express},
  volume={14},
  number={4},
  pages={1777--1799},
  year={2023},
  publisher={Optica Publishing Group}
}

@article{wang2023adaptive,
  title={Adaptive machine learning method for photoacoustic computed tomography based on sparse array sensor data},
  author={Wang, Ruofan and Zhu, Jing and Meng, Yuqian and Wang, Xuanhao and Chen, Ruimin and Wang, Kaiyue and Li, Chiye and Shi, Junhui},
  journal={Computer Methods and Programs in Biomedicine},
  volume={242},
  pages={107822},
  year={2023},
  publisher={Elsevier}
}

@article{hauptmann2020deep,
  title={Deep learning in photoacoustic tomography: current approaches and future directions},
  author={Hauptmann, Andreas and Cox, Ben},
  journal={Journal of Biomedical Optics},
  volume={25},
  number={11},
  pages={112903},
  year={2020}
}

@misc{code,
    author = "Panpan Chen and Seonyeong Park and Gangwon Jeong and Refik Mert Cam and Umberto Villa and Mark A. Anastasio",
    title  = "Benchmarking Framework for {2D} {PACT} Image Reconstruction",
    year   = "2025", 
    howpublished = "[Online]. Available from: [link provided upon acceptance\url{}]"}

@misc{data,
    author = "Panpan Chen and Seonyeong Park and Umberto Villa and Mark A. Anastasio",
    title  = "{2D} Photoacoustic Numerical Breast Phantoms and Simulated {PACT} Measurement Data",
    year   = "2025", 
    howpublished = "[Online]. Available from: [link provided upon acceptance\url{}]"}

@inproceedings{hopp2012breast,
  title={Breast tissue characterization by sound speed: Correlation with mammograms using a 2d/3d image registration},
  author={Hopp, Torsten and Ruiter, Nicole V and Duric, Neb},
  booktitle={2012 IEEE International Ultrasonics Symposium},
  pages={1--4},
  year={2012},
  organization={IEEE}
}

@article{park2025virtual,
  title={A Virtual Imaging Framework for Three-Dimensional Quantitative Optoacoustic Tomography Using Stochastic Numerical Breast Phantoms},
  author={Park, Seonyeong and Jeong, Gangwon and Villa, Umberto and Anastasio, Mark A},
  journal={arXiv preprint arXiv:2510.00189},
  year={2025}
}

@article{Kaipio2007,
    title={Statistical inverse problems: discretization, model reduction and inverse crimes},
    author={Kaipio, Jari and Somersalo, Erkki},
    journal={Journal of computational and applied mathematics},
    volume={198},
    number={2},
    pages={493--504},
    year={2007},
    publisher={Elsevier}
}

@article{sen2016accounting,
  title={Accounting for anatomical noise in search-capable model observers for planar nuclear imaging},
  author={Sen, Anando and Gifford, Howard C},
  journal={Journal of Medical Imaging},
  volume={3},
  number={1},
  pages={015502--015502},
  year={2016},
  publisher={Society of Photo-Optical Instrumentation Engineers}
}

@article{davoudi2019deep,
  title={Deep learning optoacoustic tomography with sparse data},
  author={Davoudi, Neda and De{\'a}n-Ben, Xos{\'e} Lu{\'\i}s and Razansky, Daniel},
  journal={Nature Machine Intelligence},
  volume={1},
  number={10},
  pages={453--460},
  year={2019},
  publisher={Nature Publishing Group UK London}
}

@article{yang2023recent,
  title={Recent advances in deep-learning-enhanced photoacoustic imaging},
  author={Yang, Jinge and Choi, Seongwook and Kim, Jiwoong and Park, Byullee and Kim, Chulhong},
  journal={Advanced Photonics Nexus},
  volume={2},
  number={5},
  pages={054001--054001},
  year={2023},
  publisher={Society of Photo-Optical Instrumentation Engineers}
}

@article{najafzadeh2020photoacoustic,
  title={Photoacoustic image improvement based on a combination of sparse coding and filtering},
  author={Najafzadeh, Ebrahim and Farnia, Parastoo and Lavasani, Saeedeh N and Basij, Maryam and Yan, Yan and Ghadiri, Hossein and Ahmadian, Alireza and Mehrmohammadi, Mohammad},
  journal={Journal of biomedical optics},
  volume={25},
  number={10},
  pages={106001--106001},
  year={2020},
  publisher={Society of Photo-Optical Instrumentation Engineers}
}

@article{mitcham2015photoacoustic,
  title={Photoacoustic imaging driven by an interstitial irradiation source},
  author={Mitcham, Trevor and Dextraze, Katherine and Taghavi, Houra and Melancon, Marites and Bouchard, Richard},
  journal={Photoacoustics},
  volume={3},
  number={2},
  pages={45--54},
  year={2015},
  publisher={Elsevier}
}

@article{chakraborty2023rjafroc,
  title={RJafroc: artificial intelligence systems and observer performance},
  author={Chakraborty, Dev and Zhai, Xuetong},
  journal={R package version},
  volume={2},
  number={3},
  year={2023}
}

@article{barr2013probably,
  title={Probably benign lesions at screening breast US in a population with elevated risk: prevalence and rate of malignancy in the ACRIN 6666 trial},
  author={Barr, Richard G and Zhang, Zheng and Cormack, Jean B and Mendelson, Ellen B and Berg, Wendie A},
  journal={Radiology},
  volume={269},
  number={3},
  pages={701--712},
  year={2013},
  publisher={Radiological Society of North America}
}

@inproceedings{deng2009imagenet,
  title={Imagenet: A large-scale hierarchical image database},
  author={Deng, Jia and Dong, Wei and Socher, Richard and Li, Li-Jia and Li, Kai and Fei-Fei, Li},
  booktitle={2009 IEEE conference on computer vision and pattern recognition},
  pages={248--255},
  year={2009},
  organization={Ieee}
}

@article{matthews2018parameterized,
  title={Parameterized joint reconstruction of the initial pressure and sound speed distributions for photoacoustic computed tomography},
  author={Matthews, Thomas P and Poudel, Joemini and Li, Lei and Wang, Lihong V and Anastasio, Mark A},
  journal={SIAM journal on imaging sciences},
  volume={11},
  number={2},
  pages={1560--1588},
  year={2018},
  publisher={SIAM}
}

@inproceedings{gifford2017visual,
  title={Visual-search models for location-known detection tasks},
  author={Gifford, HC and Karbaschi, Z and Banerjee, K and Das, M},
  booktitle={Medical Imaging 2017: Image Perception, Observer Performance, and Technology Assessment},
  volume={10136},
  pages={282--287},
  year={2017},
  organization={SPIE}
}

@article{tarvainen2012reconstructing,
  title={Reconstructing absorption and scattering distributions in quantitative photoacoustic tomography},
  author={Tarvainen, Tanja and Cox, Benjamin T and Kaipio, JP and Arridge, Simon R},
  journal={Inverse Problems},
  volume={28},
  number={8},
  pages={084009},
  year={2012},
  publisher={IOP Publishing}
}

@article{choe2009differentiation,
  title={Differentiation of benign and malignant breast tumors by in-vivo three-dimensional parallel-plate diffuse optical tomography},
  author={Choe, Regine and Konecky, Soren D and Corlu, Alper and Lee, Kijoon and Durduran, Turgut and Busch, David R and Pathak, Saurav and Czerniecki, Brian J and Tchou, Julia and Fraker, Douglas L and others},
  journal={Journal of biomedical optics},
  volume={14},
  number={2},
  pages={024020--024020},
  year={2009},
  publisher={Society of Photo-Optical Instrumentation Engineers}
}

@inproceedings{cam2025numerical,
  title={Numerical mouse phantoms for multispectral dynamic contrast-enhanced photoacoustic computed tomography},
  author={Cam, Refik Mert and Huang, Hsuan-Kai and Lozenski, Luke and Park, Seonyeong and Anastasio, Mark A and Villa, Umberto},
  booktitle={Photons Plus Ultrasound: Imaging and Sensing 2025},
  volume={13319},
  pages={49--54},
  year={2025},
  organization={SPIE}
}

@book{barrett2013foundations,
  title={Foundations of image science},
  author={Barrett, Harrison H and Myers, Kyle J},
  year={2013},
  publisher={John Wiley \& Sons}
}

@article{vaishnav2014objective,
  title={Objective assessment of image quality and dose reduction in CT iterative reconstruction},
  author={Vaishnav, JY and Jung, WC and Popescu, LM and Zeng, R and Myers, KJ},
  journal={Medical physics},
  volume={41},
  number={7},
  pages={071904},
  year={2014},
  publisher={Wiley Online Library}
}

@article{plativsa2011channelized,
  title={Channelized Hotelling observers for the assessment of volumetric imaging data sets},
  author={Plati{\v{s}}a, Ljiljana and Goossens, Bart and Vansteenkiste, Ewout and Park, Subok and Gallas, Brandon D and Badano, Aldo and Philips, Wilfried},
  journal={Journal of the Optical Society of America A},
  volume={28},
  number={6},
  pages={1145--1163},
  year={2011},
  publisher={Optical Society of America}
}

@article{gallas2003validating,
  title={Validating the use of channels to estimate the ideal linear observer},
  author={Gallas, Brandon D and Barrett, Harrison H},
  journal={Journal of the Optical Society of America A},
  volume={20},
  number={9},
  pages={1725--1738},
  year={2003},
  publisher={Optical Society of America}
}

@article{li2024application,
  title={Application of learned ideal observers for estimating task-based performance bounds for computed imaging systems},
  author={Li, Kaiyan and Villa, Umberto and Li, Hua and Anastasio, Mark A},
  journal={Journal of Medical Imaging},
  volume={11},
  number={2},
  pages={026002--026002},
  year={2024},
  publisher={Society of Photo-Optical Instrumentation Engineers}
}

@inproceedings{eckstein2001model,
  title={Model observers for signal-known-statistically tasks (SKS)},
  author={Eckstein, Miguel P and Abbey, Craig K},
  booktitle={Medical Imaging 2001: Image Perception and Performance},
  volume={4324},
  pages={91--102},
  year={2001},
  organization={SPIE}
}

@book{yendiki2005analysis,
  title={Analysis of signal detectability in statistically reconstructed tomographic images},
  author={Yendiki, Anastasia},
  year={2005},
  publisher={University of Michigan}
}

@inproceedings{li2023estimating,
  title={Estimating task-based performance bounds for image reconstruction methods by use of learned-ideal observers},
  author={Li, Kaiyan and Zhou, Weimin and Li, Hua and Anastasio, Mark A},
  booktitle={Medical Imaging 2023: Image Perception, Observer Performance, and Technology Assessment},
  volume={12467},
  pages={120--125},
  year={2023},
  organization={SPIE}
}

@article{jha2021objective,
  title={Objective task-based evaluation of artificial intelligence-based medical imaging methods: framework, strategies, and role of the physician},
  author={Jha, Abhinav K and Myers, Kyle J and Obuchowski, Nancy A and Liu, Ziping and Rahman, Md Ashequr and Saboury, Babak and Rahmim, Arman and Siegel, Barry A},
  journal={PET clinics},
  volume={16},
  number={4},
  pages={493--511},
  year={2021},
  publisher={Elsevier}
}

@article{petschke2013comparison,
  title={Comparison of photoacoustic image reconstruction algorithms using the channelized Hotelling observer},
  author={Petschke, Adam and La Rivi{\`e}re, Patrick J},
  journal={Journal of biomedical optics},
  volume={18},
  number={2},
  pages={026009--026009},
  year={2013},
  publisher={Society of Photo-Optical Instrumentation Engineers}
}

@inproceedings{lou2016application,
  title={Application of signal detection theory to assess optoacoustic imaging systems},
  author={Lou, Yang and Oraevsky, Alexander and Anastasio, Mark A},
  booktitle={Photons Plus Ultrasound: Imaging and Sensing 2016},
  volume={9708},
  pages={688--696},
  year={2016},
  organization={SPIE}
}

@article{huang2013full,
  title={Full-wave iterative image reconstruction in photoacoustic tomography with acoustically inhomogeneous media},
  author={Huang, Chao and Wang, Kun and Nie, Liming and Wang, Lihong V and Anastasio, Mark A},
  journal={IEEE transactions on medical imaging},
  volume={32},
  number={6},
  pages={1097--1110},
  year={2013},
  publisher={IEEE}
}

@article{chen2025learning,
  title={Learning a Filtered Backprojection Reconstruction Method for Photoacoustic Computed Tomography with Hemispherical Measurement Geometries},
  author={Chen, Panpan and Park, Seonyeong and Cam, Refik Mert and Huang, Hsuan-Kai and Oraevsky, Alexander A and Villa, Umberto and Anastasio, Mark A},
  journal={IEEE Transactions on Medical Imaging},
  year={2025},
  publisher={IEEE}
}

@article{pattyn2021model,
  title={Model-based optical and acoustical compensation for photoacoustic tomography of heterogeneous mediums},
  author={Pattyn, Alexander and Mumm, Zackary and Alijabbari, Naser and Duric, Neb and Anastasio, Mark A and Mehrmohammadi, Mohammad},
  journal={Photoacoustics},
  volume={23},
  pages={100275},
  year={2021},
  publisher={Elsevier}
}

@article{mitsuhashi2014investigation,
  title={Investigation of the far-field approximation for modeling a transducer's spatial impulse response in photoacoustic computed tomography},
  author={Mitsuhashi, Kenji and Wang, Kun and Anastasio, Mark A},
  journal={Photoacoustics},
  volume={2},
  number={1},
  pages={21--32},
  year={2014},
  publisher={Elsevier}
}

@article{poudel2019survey,
  title={A survey of computational frameworks for solving the acoustic inverse problem in three-dimensional photoacoustic computed tomography},
  author={Poudel, Joemini and Lou, Yang and Anastasio, Mark A},
  journal={Physics in Medicine \& Biology},
  volume={64},
  number={14},
  pages={14TR01},
  year={2019},
  publisher={IOP Publishing}
}

@inproceedings{schwab2019learned,
  title={Learned backprojection for sparse and limited view photoacoustic tomography},
  author={Schwab, Johannes and Antholzer, Stephan and Haltmeier, Markus},
  booktitle={Photons Plus Ultrasound: Imaging and Sensing 2019},
  volume={10878},
  pages={263--271},
  year={2019},
  organization={SPIE}
}

@article{dehner2023deep,
  title={A deep neural network for real-time optoacoustic image reconstruction with adjustable speed of sound},
  author={Dehner, Christoph and Zahnd, Guillaume and Ntziachristos, Vasilis and J{\"u}stel, Dominik},
  journal={Nature Machine Intelligence},
  volume={5},
  number={10},
  pages={1130--1141},
  year={2023},
  publisher={Nature Publishing Group UK London}
}

@article{chen2020improved,
  title={Improved photoacoustic imaging of numerical bone model based on attention block {U}-net deep learning network},
  author={Chen, Panpan and Liu, Chengcheng and Feng, Ting and Li, Yong and Ta, Dean},
  journal={Applied Sciences},
  volume={10},
  number={22},
  pages={8089},
  year={2020},
  publisher={MDPI}
}

@inproceedings{jeon2020deep,
  title={Deep learning-based speed of sound aberration correction in photoacoustic images},
  author={Jeon, Seungwan and Kim, Chulhong},
  booktitle={Photons plus ultrasound: Imaging and sensing 2020},
  volume={11240},
  pages={24--27},
  year={2020},
  organization={SPIE}
}

@article{hauptmann2018model,
  title={Model-based learning for accelerated, limited-view {3-D} photoacoustic tomography},
  author={Hauptmann, Andreas and Lucka, Felix and Betcke, Marta and Huynh, Nam and Adler, Jonas and Cox, Ben and Beard, Paul and Ourselin, Sebastien and Arridge, Simon},
  journal={IEEE transactions on medical imaging},
  volume={37},
  number={6},
  pages={1382--1393},
  year={2018},
  publisher={IEEE}
}

@article{yao2016multiscale,
  title={Multiscale functional and molecular photoacoustic tomography},
  author={Yao, Junjie and Xia, Jun and Wang, Lihong V},
  journal={Ultrasonic Imaging},
  volume={38},
  number={1},
  pages={44--62},
  year={2016},
  publisher={SAGE Publications Sage CA: Los Angeles, CA}
}

@article{xia2014photoacoustic,
  title={Photoacoustic tomography: principles and advances},
  author={Xia, Jun and Yao, Junjie and Wang, Lihong V},
  journal={Electromagnetic waves (Cambridge, Mass.)},
  volume={147},
  pages={1},
  year={2014},
  publisher={NIH Public Access}
}

@article{wang2012simple,
  title={A simple {F}ourier transform-based reconstruction formula for photoacoustic computed tomography with a circular or spherical measurement geometry},
  author={Wang, Kun and Anastasio, Mark A},
  journal={Physics in Medicine \& Biology},
  volume={57},
  number={23},
  pages={N493},
  year={2012},
  publisher={IOP Publishing}
}

@article{xu2004time,
  title={Time reversal and its application to tomography with diffracting sources},
  author={Xu, Yuan and Wang, Lihong V},
  journal={Physical review letters},
  volume={92},
  number={3},
  pages={033902},
  year={2004},
  publisher={APS}
}

@article{frikel2015artifacts,
  title={Artifacts in incomplete data tomography with applications to photoacoustic tomography and sonar},
  author={Frikel, Jürgen and Quinto, Eric Todd},
  journal={SIAM Journal on Applied Mathematics},
  volume={75},
  number={2},
  pages={703--725},
  year={2015},
  publisher={SIAM}
}

@article{kunyansky2007explicit,
  title={Explicit inversion formulae for the spherical mean {R}adon transform},
  author={Kunyansky, Leonid A},
  journal={Inverse Problems},
  volume={23},
  number={1},
  pages={373},
  year={2007},
  publisher={IOP Publishing}
}

@article{xu2005universal,
  title={Universal back-projection algorithm for photoacoustic computed tomography},
  author={Xu, Minghua and Wang, Lihong V},
  journal={Physical Review E},
  volume={71},
  number={1},
  pages={016706},
  year={2005},
  publisher={APS}
}

@article{xu2004reconstructions,
  title={Reconstructions in limited-view thermoacoustic tomography},
  author={Xu, Yuan and Wang, Lihong V and Ambartsoumian, Gaik and Kuchment, Peter},
  journal={Medical Physics},
  volume={31},
  number={4},
  pages={724--733},
  year={2004},
  publisher={Wiley Online Library}
}

@article{buehler2011model,
  title={Model-based optoacoustic inversions with incomplete projection data},
  author={Buehler, Andreas and Rosenthal, Amir and Jetzfellner, Thomas and Dima, Alexander and Razansky, Daniel and Ntziachristos, Vasilis},
  journal={Medical physics},
  volume={38},
  number={3},
  pages={1694--1704},
  year={2011},
  publisher={Wiley Online Library}
}

@inproceedings{ronneberger2015u,
  title={U-Net: Convolutional networks for biomedical image segmentation},
  author={Ronneberger, Olaf and Fischer, Philipp and Brox, Thomas},
  booktitle={Medical Image Computing and Computer-Assisted Intervention--MICCAI 2015: 18th International Conference, Munich, Germany, October 5-9, 2015, Proceedings, Part III 18},
  pages={234--241},
  year={2015},
  organization={Springer}
}

@article{fang2009monte,
  title={Monte {Carlo} simulation of photon migration in {3D} turbid media accelerated by graphics processing units},
  author={Fang, Qianqian and Boas, David A},
  journal={Optics Express},
  volume={17},
  number={22},
  pages={20178--20190},
  year={2009},
  publisher={Optica Publishing Group}
}

@article{wang2012photoacoustic,
  title={Photoacoustic tomography: in vivo imaging from organelles to organs},
  author={Wang, Lihong V and Hu, Song},
  journal={science},
  volume={335},
  number={6075},
  pages={1458--1462},
  year={2012},
  publisher={American Association for the Advancement of Science}
}

@article{grohl2021deep,
  title={Deep learning for biomedical photoacoustic imaging: A review},
  author={Gr{\"o}hl, Janek and Schellenberg, Melanie and Dreher, Kris and Maier-Hein, Lena},
  journal={Photoacoustics},
  volume={22},
  pages={100241},
  year={2021},
  publisher={Elsevier}
}

@article{jeon2021deep,
  title={A deep learning-based model that reduces speed of sound aberrations for improved in vivo photoacoustic imaging},
  author={Jeon, Seungwan and Choi, Wonseok and Park, Byullee and Kim, Chulhong},
  journal={IEEE transactions on image processing},
  volume={30},
  pages={8773--8784},
  year={2021},
  publisher={IEEE}
}

@article{hacker2024tutorial,
  title={Tutorial on phantoms for photoacoustic imaging applications},
  author={Hacker, Lina and Joseph, James and Lilaj, Ledia and Manohar, Srirang and Ivory, Aoife M and Tao, Ran and Bohndiek, Sarah E and Members of IPASC},
  journal={Journal of biomedical optics},
  volume={29},
  number={8},
  pages={080801--080801},
  year={2024},
  publisher={Society of Photo-Optical Instrumentation Engineers}
}

@article{christie2023review,
  title={Review of imaging test phantoms},
  author={Christie, Liam B and Zheng, Wenhan and Johnson, William and Marecki, Eric K and Heidrich, James and Xia, Jun and Oh, Kwang W},
  journal={Journal of Biomedical Optics},
  volume={28},
  number={8},
  pages={080903--080903},
  year={2023},
  publisher={Society of Photo-Optical Instrumentation Engineers}
}

@article{lou2017generation,
  title={Generation of anatomically realistic numerical phantoms for photoacoustic and ultrasonic breast imaging},
  author={Lou, Yang and Zhou, Weimin and Matthews, Thomas P and Appleton, Catherine M and Anastasio, Mark A},
  journal={Journal of biomedical optics},
  volume={22},
  number={4},
  pages={041015--041015},
  year={2017},
  publisher={Society of Photo-Optical Instrumentation Engineers}
}

@article{dispirito2021sounding,
  title={Sounding out the hidden data: a concise review of deep learning in photoacoustic imaging},
  author={DiSpirito III, Anthony and Vu, Tri and Pramanik, Manojit and Yao, Junjie},
  journal={Experimental Biology and Medicine},
  volume={246},
  number={12},
  pages={1355--1367},
  year={2021},
  publisher={SAGE Publications Sage UK: London, England}
}

@article{huang2021functional,
  title={Functional multispectral optoacoustic tomography imaging of hepatic steatosis development in mice},
  author={Huang, Shan and Blutke, Andreas and Feuchtinger, Annette and Klemm, Uwe and Zachariah Tom, Robby and Hofmann, Susanna M and Stiel, Andre C and Ntziachristos, Vasilis},
  journal={EMBO Molecular Medicine},
  volume={13},
  number={9},
  pages={e13490},
  year={2021}
}

@article{ozdemir2022oadat,
  title={OADAT: Experimental and synthetic clinical optoacoustic data for standardized image processing},
  author={Ozdemir, Firat and Lafci, Berkan and Dean-Ben, Xose Luis and Razansky, Daniel and Perez-Cruz, Fernando},
  journal={arXiv preprint arXiv:2206.08612},
  year={2022}
}

@article{park2023stochastic,
  title={Stochastic three-dimensional numerical phantoms to enable computational studies in quantitative optoacoustic computed tomography of breast cancer},
  author={Park, Seonyeong and Villa, Umberto and Li, Fu and Cam, Refik Mert and Oraevsky, Alexander A and Anastasio, Mark A},
  journal={Journal of biomedical optics},
  volume={28},
  number={6},
  pages={066002--066002},
  year={2023},
  publisher={Society of Photo-Optical Instrumentation Engineers}
}

@article{bench2020toward,
  title={Toward accurate quantitative photoacoustic imaging: learning vascular blood oxygen saturation in three dimensions},
  author={Bench, Ciaran and Hauptmann, Andreas and Cox, Ben},
  journal={Journal of Biomedical Optics},
  volume={25},
  number={8},
  pages={085003--085003},
  year={2020},
  publisher={Society of Photo-Optical Instrumentation Engineers}
}

@article{hamarneh2010vascusynth,
  title={VascuSynth: Simulating vascular trees for generating volumetric image data with ground-truth segmentation and tree analysis},
  author={Hamarneh, Ghassan and Jassi, Preet},
  journal={Computerized medical imaging and graphics},
  volume={34},
  number={8},
  pages={605--616},
  year={2010},
  publisher={Elsevier}
}

@article{li20213,
  title={3-D stochastic numerical breast phantoms for enabling virtual imaging trials of ultrasound computed tomography},
  author={Li, Fu and Villa, Umberto and Park, Seonyeong and Anastasio, Mark A},
  journal={IEEE transactions on ultrasonics, ferroelectrics, and frequency control},
  volume={69},
  number={1},
  pages={135--146},
  year={2021},
  publisher={IEEE}
}

@article{staal2004ridge,
  title={Ridge-based vessel segmentation in color images of the retina},
  author={Staal, Joes and Abr{\`a}moff, Michael D and Niemeijer, Meindert and Viergever, Max A and Van Ginneken, Bram},
  journal={IEEE transactions on medical imaging},
  volume={23},
  number={4},
  pages={501--509},
  year={2004},
  publisher={IEEE}
}

@article{lin2018single,
  title={Single-breath-hold photoacoustic computed tomography of the breast},
  author={Lin, Li and Hu, Peng and Shi, Junhui and Appleton, Catherine M and Maslov, Konstantin and Li, Lei and Zhang, Ruiying and Wang, Lihong V},
  journal={Nature communications},
  volume={9},
  number={1},
  pages={2352},
  year={2018},
  publisher={Nature Publishing Group UK London}
}

@article{treeby2010k,
  title={k-Wave: MATLAB toolbox for the simulation and reconstruction of photoacoustic wave fields},
  author={Treeby, Bradley E and Cox, Benjamin T},
  journal={Journal of biomedical optics},
  volume={15},
  number={2},
  pages={021314--021314},
  year={2010},
  publisher={Society of Photo-Optical Instrumentation Engineers}
}

@article{sickles2013acr,
  title={ACR BI-RADS{\textregistered} Atlas, Breast imaging reporting and data system.},
  author={Sickles, Edward A},
  journal={American College of Radiology.},
  pages={39},
  year={2013}
}

@article{jeong2025revisiting,
  title={Revisiting the joint estimation of initial pressure and speed-of-sound distributions in photoacoustic computed tomography with consideration of canonical object constraints},
  author={Jeong, Gangwon and Villa, Umberto and Anastasio, Mark A},
  journal={Photoacoustics},
  volume={43},
  pages={100700},
  year={2025},
  publisher={Elsevier}
}

@article{beck2009fast,
  title={A fast iterative shrinkage-thresholding algorithm for linear inverse problems},
  author={Beck, Amir and Teboulle, Marc},
  journal={SIAM journal on imaging sciences},
  volume={2},
  number={1},
  pages={183--202},
  year={2009},
  publisher={SIAM}
}

@article{wang2004image,
  title={Image quality assessment: from error visibility to structural similarity},
  author={Wang, Zhou and Bovik, Alan C and Sheikh, Hamid R and Simoncelli, Eero P},
  journal={IEEE transactions on image processing},
  volume={13},
  number={4},
  pages={600--612},
  year={2004},
  publisher={IEEE}
}
\bibliographystyle{spiejour}  

%%%%% Biographies of authors %%%%%
\vspace{2ex}\noindent\textbf{Panpan Chen} is a Ph.D. candidate in the Department of Bioengineering at the University of Illinois Urbana–Champaign, Urbana, IL, USA. She received the B.S. degree in Optoelectronic Information Science and Engineering from Anhui Normal University, Wuhu, China, in 2018, and the M.S. degree in Acoustics from Tongji University, Shanghai, China, in 2022. Her research interests include medical imaging, deep learning and image quality assessment. She is a 2024–2025 Mavis Future Faculty Fellow at the Grainger College of Engineering, a 2025 McGinnis Medical Innovation Graduate Fellow in the Department of Bioengineering, and a student member of SPIE.

\vspace{2ex}\noindent\textbf{Seonyeong Park} received the B.S. degree in Electronics, Computer, and Telecommunication Engineering and the M.S. degree in Information and Communications Engineering from Pukyong National University, Busan, Korea, in 2011 and 2013, respectively. She earned the Ph.D. degree in Electrical and Computer Engineering from Virginia Commonwealth University, Richmond, VA, USA, in 2017. She is currently a Research Assistant Professor in the Department of Bioengineering at the University of Illinois Urbana-Champaign, Urbana, IL, USA. Her research interests include photoacoustic computed tomography, numerical modeling of biomedical objects, image reconstruction, and inverse problems. 

\vspace{2ex}\noindent\textbf{Gangwon Jeong} is currently a PhD student in the Department of Bioengineering at the University of Illinois at Urbana-Champaign, Urbana, IL, USA. He received his B.S. and M.S. degrees in Energy Resources Engineering and Energy Systems Engineering from Seoul National University in 2014 and 2016, respectively. His research focuses on medical image reconstruction, particularly in ultrasound and photoacoustic computed tomography, and the application of machine learning in medical imaging. 

\vspace{2ex}\noindent\textbf{Refik Mert Cam} is a Ph.D. candidate in Electrical and Computer Engineering at the University of Illinois Urbana-Champaign. He received his B.S. degree in Electrical and Electronics Engineering from Middle East Technical University, Turkey, in 2020. His research interests broadly lie in signal processing and deep learning, with a particular focus on imaging applications. Refik is a 2023–2024 Mavis Future Faculty Fellow at the Grainger College of Engineering and a student member of IEEE and SPIE. 

\vspace{2ex}\noindent\textbf{Umberto Villa} received the B.S. and M.S. degrees in Mathematical Engineering from the Politecnico di Milano, Milan, Italy, in 2005 and 2007, respectively, and the Ph.D. degree in Mathematics from Emory University, Atlanta, GA, USA, in 2012. He is an Assistant Professor at the Department of Biomedical Engineering and Core Faculty of the Oden Institute for Computational Engineering and Science, The University of Texas at Austin, Austin, TX, USA. His research interests lie in the computational and mathematical aspects of large-scale inverse problems, imaging science, and uncertainty quantification. 

\vspace{2ex}\noindent\textbf{Mark A. Anastasio} received the Ph.D. degree from The University of Chicago, Chicago, IL, USA, in 2001. He is currently a Donald Biggar Willett Professor of Engineering and the Head of the Department of Bioengineering, University of Illinois at Urbana–Champaign, Urbana, IL, USA. His research broadly addresses computational image science, inverse problems in imaging, and machine learning for imaging applications. He has contributed broadly to emerging biomedical imaging technologies that include photoacoustic computed tomography, ultrasound computed tomography, and X-ray phase-contrast imaging. His work has been supported by numerous NIH grants and he served for two years as the Chair of NIH EITA Study Section. Dr. Anastasio is a fellow of the Society of Photo-Optical Instrumentation Engineers (SPIE), American Institute for Medical and Biological Engineering (AIMBE), the International Academy of Medical and Biological Engineering (IAMBE), and the Institute of Electrical and Electronics Engineers (IEEE). \\

\listoffigures
\listoftables

\end{spacing}
\end{document}